\documentstyle[11pt,aaspp4,flushrt,epsf]{article}

\newcommand{\etal}{{\it et al.\ }}
\newcommand{\eg}{{\it e.g.,\ }}

\newcommand{\ltsima}{$\; \buildrel < \over \sim \;$}
\newcommand{\lsim}{\lower.5ex\hbox{\ltsima}}
\newcommand{\gtsima}{$\; \buildrel > \over \sim \;$}
\newcommand{\gsim}{\lower.5ex\hbox{\gtsima}}

\begin{document}

\journalid{337}{25 January 1998}
\articleid{11}{14}

\title{Clustering at High Redshift:
Precise Constraints from a Deep, Wide Area Survey}

\author{Marc Postman\altaffilmark{1}}
\affil{Space Telescope Science Institute\altaffilmark{2},
Baltimore, MD 21218}
 
\author{Tod R. Lauer}
\affil{National Optical Astronomy Observatories\altaffilmark{3},
Tucson, AZ 85726}
 
\author{Istv\'an Szapudi}
\affil{University of Durham, Department of Physics,
South Road, Durham, DH1 3LE, UK}
 
\author{William Oegerle\altaffilmark{1}}
\affil{Johns Hopkins University, Department of Physics \& Astronomy,
Baltimore, MD 21218}
 
\vskip 1 cm
\centerline{Accepted for publication in the {\it Astrophysical Journal}}

\altaffiltext{1}{Visiting Astronomer Kitt Peak National Observatory,
NOAO.}
\altaffiltext{2}{The Space Telescope Science Institute
is operated by the Association of Universities for
Research in Astronomy (AURA), Inc., under National Aeronautics and
Space Administration (NASA) Contract NAS 5-26555.}
\altaffiltext{3}{The National Optical Astronomy Observatories are
operated by AURA, Inc., under cooperative agreement with the National
Science Foundation.}

\newpage
\begin{abstract}
We present constraints on the evolution of large-scale structure
from a catalog of 710,000 galaxies with $I_{AB} \le 24$
derived from a KPNO 4m CCD imaging survey of a contiguous
$4^{\circ} \times 4^{\circ}$ region.
The advantage of using large contiguous surveys for measuring
clustering properties on even modest angular scales is substantial:
the effects of cosmic scatter are strongly suppressed.
We provide highly accurate measurements
of the two-point angular correlation function, $\omega(\theta)$, as a function
of magnitude on scales up to $1.5^{\circ}$. 
The amplitude of $\omega(\theta)$ declines by a factor of 
$\sim 10$ over the range $16 \le I \le 20$ but only by a factor
of $2 - 3$ over the range $20 < I \le 23$.  
For a redshift dependence of the spatial correlation function, $\xi(r)$, 
parameterized as $\xi(r,z) = ({r \over r_o})^{-\gamma} 
(1 + z)^{-(3+\epsilon)}$, we find $r_o = 5.2\pm0.4 h^{-1}$ Mpc,
and $\epsilon \gsim 0$ for $I \le 20$. This is in good
agreement with the results from local redshift surveys.
At $I > 20$, our best fit values shift towards lower $r_o$ and
more negative $\epsilon$. A strong covariance between $r_o$ and
$\epsilon$ prevent us from rejecting $\epsilon > 0$ even at faint
magnitudes but if $\epsilon > 1$, we strongly reject
$r_o \lsim 4h^{-1}$ Mpc (co-moving). The above expression for
$\xi(r,z)$ and our data give a correlation length 
of $r_o(z = 0.5) \approx 3.0\pm0.4h^{-1}$ Mpc, about
a factor of 2 larger than the correlation length at $z = 0.5$ 
derived from the Canada--France Redshift Survey (CFRS; \cite{lefev96}). 
The small volume sampled by the CFRS and other
deep redshift probes, however, make these spatial surveys strongly
susceptible to cosmic scatter and will tend to bias their derived
correlation lengths low.
Our results are consistent with redshift distributions in which
$\sim30-50$\% of the galaxies at $I = 23$ lie at $z > 1$.
The best fit power law slope of the correlation function remains
independent of $I$ magnitude for $I \le 22$. At fainter limits,
there is a suggestive trend towards flatter slopes that occurs
at fluxes consistent with similar trends seen by \cite{nwind} and \cite{campos}.
The galaxy counts span 11 magnitudes and provide an 
accurate calibration of the galaxy surface density. 
We find evidence for mild galaxy evolution -- about 1 mag of
brightening or a doubling of the density by $I=23$ relative
to an $\Omega_o = 1$ no evolution model; about 0.5 mag of brightening
or a factor of 1.5 increase in surface density relative to an
open model. 
Our galaxy counts agree well with those from the HDF survey and, thus,
argue against a significant inclusion of sub-galactic components in
the latter census for $I < 24$. 

\end{abstract}

\keywords{large-scale structure, clustering, galaxy evolution, galaxy 
catalogs}

\section{Introduction}

The evolution of large-scale structure (LSS) in the universe probes
the abundance and form of dark matter,
the mean baryon density, the turnover scale in the perturbation power
spectrum, and the formation processes of galaxies and clusters.
Observational constraints on the evolution of LSS
provide strong limits on structure formation theories because
the sensitivity of the mass function to $\Omega_{\circ}$ is high and because
coherent structures on scales $>10h^{-1}$ Mpc are still in the linear regime,
precisely where important distinctions can be made between competing theories.
A study of this evolution is, thus, fundamental.
Substantial progress has been made in studying LSS evolution from 
deep redshift surveys. The general consensus from the most distant probes
(\cite{cohen}, \cite{steidel}, \cite{mauro}) suggests
that there has been significant
evolution in the clustering properties of galaxies since $z \sim 3$. 
Precisely just how much evolution has occurred since $z \sim 1$  
is less certain but the data are being accumulated
rapidly (\cite{CFRS}, \cite{lefev}, \cite{connolly}, \cite{ajc98}). 

Two-dimensional surveys offer some key advantages
over narrow, deep redshift surveys like those above. Specifically,
a large area angular survey subtends many galaxy correlation lengths in
the transverse direction enabling many independent cells to contribute
to the signal at a given depth. 
In addition, the power spectra derived from angular surveys
are unaffected by redshift distortions, which tend to wash out
small-scale power. The disadvantage of a 2D survey, of course, is that
the depth sampled in a given flux range is usually broad and one must
therefore deal with projection effects. 
Observationally, mapping large areas of sky is
now quite an efficient process due to the availability of large-format
CCD arrays and mosaic cameras. Ideally, one would like to conduct
deep redshift surveys that also cover large, quasi-contiguous areas.
As an initial step towards this goal,
we have surveyed a contiguous 16 square degree area
in the $I$-band using the prime focus CCD camera on the Mayall 4m at
Kitt Peak National Observatory. The area is about an order
of magnitude larger than any comparably deep, contiguous survey yet 
published. Our goal is to set accurate limits at $z = 1$
on the amplitude of structures now seen at $z = 0$.

The study of large-scale structure has advanced primarily through the
study of the moments of the galaxy distribution. These fundamental functions,
which must be explained by any viable structure formation model, are
determined, in part, by the initial conditions at recombination
and by subsequent non-linear growth. The presence of non-zero high order
moments ($n \ge 3$) provide constraints on the non-Gaussian properties of the 
galaxy distribution and, hence, provide very strong discrimination between
competing models. The first two moments, however, are essential
for assessing departures from a non-evolving, Poisson distribution.
The combination of area and depth provided by this survey
($\sim 710,000$ galaxies with $I_{AB} \le 24$)
allows us to determine the first two moments
with significantly lower uncertainties than previous works. 
For example, the angular two-point correlation function at faint
magnitudes has been measured on scales of $\lsim 10$ arcminutes 
(\eg \cite{ef91}, \cite{nwind}, \cite{campos}, \cite{lp96}, \cite{wf97}, 
\cite{bs97}) but there is substantial scatter in the results at $I \gsim 21$, 
which has lead to conflicting conclusions about the evolution of structure. 
A significant reason for the scatter is the relatively small 
($\lsim 0.25$ deg$^2$) contiguous areas used. At these scales, one is subject 
to both cosmic variations in clustering and systematic corrections that
are comparable in amplitude with the signal being measured. Such effects 
are negligible for the present survey and thus enable a major enhancement
to the precision and accuracy of LSS and evolution measurements. 
Furthermore, the survey supports a reliable determination of at least
the 3rd and 4th moments of the galaxy distribution.

In this paper we present the $I$-band galaxy counts and constraints on
the evolution of the two-point correlation
function. An analysis of the higher moments ($n = 3,4$) of the
galaxy distribution will be presented in a separate paper (\cite{npt98}).
Section 2 contains a brief description of the observations and survey
strategy. Section 3 presents results for the $I$-band galaxy counts.
Our results for the two-point angular correlation function are presented
in Section 4. The interpretation of our results is discussed in Section 5.
We adopt $h = H_o/$(100 km sec$^{-1}$ Mpc$^{-1}$). 

\section{Survey Strategy and Observations}

The primary goal of the survey is to use distant clusters of galaxies 
to explore LSS out to $z = 1$. To do this reliably, the survey had to subtend 
at least $\sim75h^{-1}$ Mpc in at least one of the angular directions given
that the local correlation length of clusters is $15 - 20h^{-1}$ Mpc 
(\cite{phg92}, \cite{dalton94}, \cite{croft97}).
At $z = 1$, 75$h^{-1}$ Mpc subtends $4.1^{\circ}$ for $q_o = 0.1$ 
($4.9^{\circ}$ for $q_o = 0.5$). Furthermore,
to measure the high redshift cluster correlation function with accuracies
approaching those of low redshift studies, 
the survey needed to contain at least 150 -- 200 clusters 
with $z \geq 0.3$.  Estimates of the surface density of clusters
(\cite{gunn}, \cite{p96}) are in the range 10 -- 15 clusters/deg$^2$ out
to $z = 1$, thus an area of at least 15 square degrees was needed.
Additional survey size constraints were imposed by the desire
to accurately measure the galaxy angular two-point correlation function,
$\omega(\theta)$, on scales up to $\sim 1^{\circ}$ at
faint magnitudes. This required that the survey subtend at least
$3^{\circ}$ to minimize the effects of the integral constraint on
$\omega(\theta)$ on scales $\theta < 1^{\circ}$.
These requirements, plus the desire to avoid severe aliasing effects and 
unfavorable window functions, led us to choose a 
$4^{\circ} \times 4^{\circ}$ survey geometry.

The KPNO 4m prime focus CCD camera has a $16'$ field 
of view (0.47 arcseconds per pixel); 256 exposures were, therefore, required
to survey the entire field. 
Each pointing overlapped its adjacent pointing by 1 arcminute. 
Local stability of the zeropoint is provided by comparing the overlap
regions of any given CCD field with its neighbors.
By comparing the photometry of objects in 
common in any overlap region, we find that
we can measure the relative photometric offsets between any two images
to 0.03 mag.  We then solve for relative offsets of all images in the
survey by a system of linear equations that minimizes the local
differences between the images and their neighbors.
Simulations of this operation show that the local error in the
zeropoint of any image is $\sim0.016$ mag which translates
to a limit on the random error in 
$\omega(\theta)$ of $\lsim 0.001$ 
due to frame-to-frame zeropoint variations.
As roughly half of the images were obtained in non-photometric
conditions, this allows us to still calibrate the survey
from the photometric set of images.
Unfortunately, building the calibration from a set of local differences
is poor at preserving a constant zeropoint over large scales.
To provide for stability of the zeropoint over the extent of the
survey, we used large-area calibration frames obtained
under photometric conditions at the KPNO 0.9m telescope by Ralph Shupping.
The calibration frames were interspersed throughout the survey
area, and provide a check against large-scale gradients in the calibration.
Based on the quality of this material, we estimate that the zeropoint
is constant over both $4^\circ$ dimensions of the survey to $\lsim 0.04$ mag
which translates to a systematic error in $\omega(\theta) \lsim 0.003$
on a $4^{\circ}$ scale and proportionally less on smaller scales. 
Indeed, computation of $\omega(\theta)$ in the independent quadrants of the 
survey yields results which are consistent with one another at the above 
level.

We selected the field centered at
10h~13m~27.95s  +52d~36m~43.5s (J2000)
by virtue of its high galactic latitude ($+51^{\circ}$),
low HI column density ($2.2 \times 10^{20}\ {\rm cm}^{-2}$),
high declination (extended visibility from KPNO), low
IRAS $100\mu$ cirrus emission,
and the absence of many bright stars or nearby rich clusters.
In a survey of this size, however, it is impossible to avoid all
bright ($I \le 16$) stars. 
The effective area of the survey is 14.7 deg$^2$ after
we exclude those pixels in the vicinity of these bright objects.

Each exposure was 900 seconds in duration yielding images with
a $5\sigma$ limit at $I_{AB}=24$ and sufficient depth to
detect cluster galaxies 2 magnitudes fainter than the typical
unevolved first-ranked elliptical at $z = 1.$
This depth was essential --- a shallower survey would only be
sufficient for detecting the very richest $z \sim 1$ clusters
(at $I_{AB} = 24$, we are able to 
detect $z \sim 1$ Abell richness class 1 systems)
and would limit visibility of structure evolution.

\section{Galaxy Counts} 

Automated object detection and classification were performed using a modified
version of the FOCAS package (\cite{valdes}). 
Modifications include the use of a position-dependent point 
spread function (PSF), an essential feature for accurate star/galaxy
classification in data obtained with 
the KPNO PFCCD system. Extensive catalog quality
assurance was performed to verify object classification accuracy
and to remove spurious detections near bright objects. A detailed description
of the galaxy catalog construction process is discussed in \cite{p98}.
However, it is important to provide a brief description of the star-galaxy 
classification procedure here. 
The object classification is done on a frame-by-frame
basis. First, the PSF in each image is determined using a set of compact, 
symmetric objects
(this is accomplished with the FOCAS {\it autopsf} routine). To assure the code
has not selected inappropriate objects, each derived PSF is 
visually inspected and a quick comparison is made with the actual image frame to
assess whether the PSF template is sensible. In about 10 -- 15\% of the cases,
stellar templates had to be selected manually. Second, a 
polynomial representation of the PSF variation as a function of distance, $r$,
from the camera's optical center
is made. (Specifically, we find the function $B(r) = a_0 + a_2 r^2 + a_4 r^4$
is an accurate model of the variation in the KPNO camera prior to the installation
of the new prime focus corrector/ADC in June 1997).
The optical center is determined directly from the data by mapping
the stellar surface density (using a position-independent PSF classifier)
as a function of CCD position. Failure to use a position-dependent PSF
results in a paucity of stars near the CCD edges which traces
the optical distortions. 
Finally, each detected object is classified as a star or galaxy based
on fitting both an unbroadened and broadened PSF to the two-dimensional object
image. Any object which requires a component broader than 1.3 times
the nominal PSF {\it at the object's position on the frame} is considered to be a
galaxy. The stellar surface density derived from position-dependent PSF
classification shows no variation with distance from the optical center.
This classification method works well down to about I=21.5 in these data (based
on running the same code on simulations). Fainter than this, a statistical
approach is needed to assess the probability that an object is a galaxy. 
However, for $I > 21.5$ galaxies outnumber stars by at least a factor of 5
(the galaxy/star ratio is $\sim10$ by $I=23.5$, based on an extrapolation
of the star and galaxy counts measured to $I=21.5$). 
Hence, star contamination would not exceed
$\sim 10 - 15$\% even if one simply classified every object fainter than
$I = 21.5$ as a galaxy. We, however, do use a probability derived from
the extrapolated galaxy/star ratio to classify objects fainter than $I=21.5$.

\begin{figure}
\epsscale{1.0}
\plotone{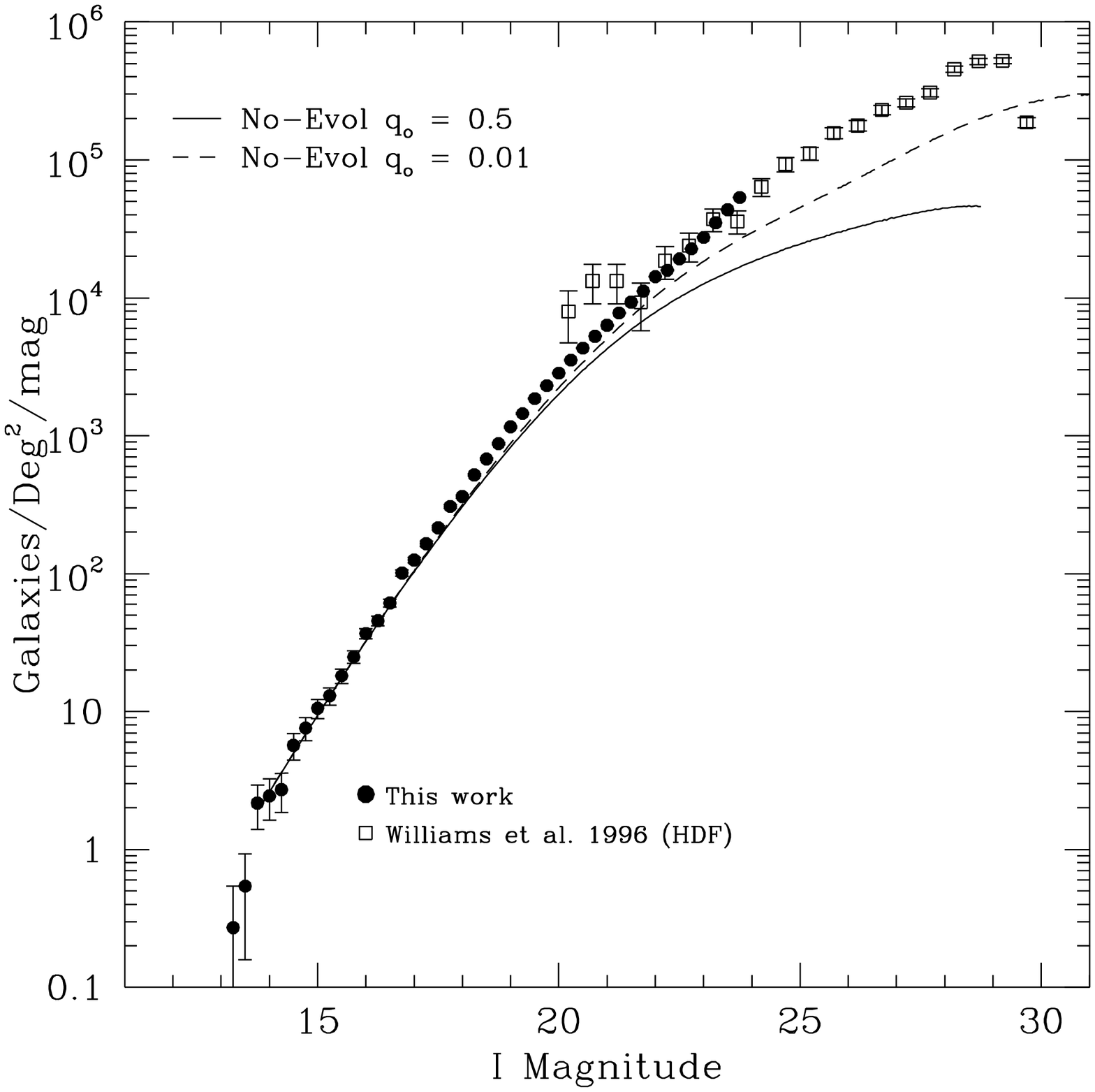}
\caption{The differential galaxy surface density
as a function of isophotal $I$ magnitude. The $I-$band counts from the
Hubble Deep Field (open circles; \cite{hdf}) are shown as well.
The models (dashed and solid lines) are from \cite{fb98}.}
\label{logNmag} 
\end{figure}

\begin{figure}
\epsscale{1.0}
\plotone{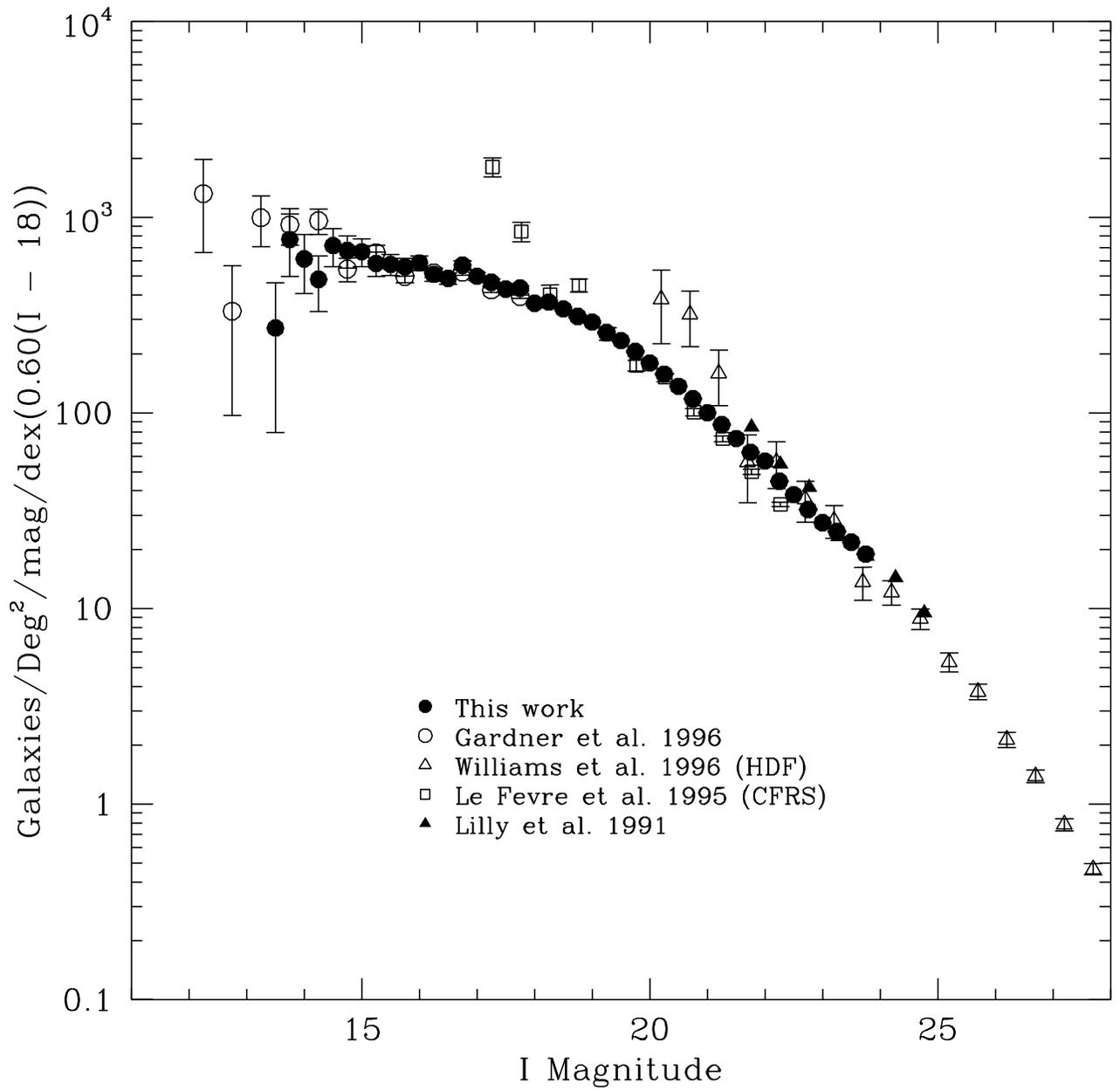}
\caption{The differential galaxy surface density
as a function of isophotal $I$ magnitude for various surveys.
Counts have been normalized by dex(0.6($I-18$)) to enhance small
variations and departures from the expectations of
Euclidean space.}
\label{compNmag}
\end{figure}

Figure~\ref{logNmag} and Table~1 show the differential number of galaxies
per unit area per magnitude as a function of $I$ magnitude.
Also shown in Figure~\ref{logNmag} are
the predictions from the dwarf-dominated, no-evolution models of \cite{fb98}
and the counts from the Hubble Deep Field (\cite{hdf}) all adjusted
to the Cousins $I$ system. 
The curvature seen in the $I-$band log N-mag relationship
is real -- the effects of object misclassification and incompleteness 
are negligible for $I \le 21$ and 
have been accurately modeled and removed at fainter magnitudes as discussed
above.  The slope of the relationship monotonically
decreases in a nearly linear fashion from the 
Euclidean value of 0.60 at $I = 15$ to 0.27 by $I = 23.5$.
Figure~\ref{compNmag} demonstrates how our counts compare with
previous determinations. For clarity, we have normalized all results
by dex$(0.6(I-18))$. 
Our counts are in good agreement with those of 
\cite{lilly91}, \cite{smail95}, 
\cite{lefev} (CFRS), \cite{hdf}, and \cite{gar96}. 
Our survey provides a homogeneous calibration of the $I-$band galaxy
counts over an 11 magnitude range
($13 \le I \le 24$), and actually covers more
area at the {\it bright} end ($I < 19$) than any of these earlier surveys.
This is an important step towards removing ambiguities in the galaxy
number counts at brighter magnitudes.
We find that the CFRS counts are systematically high for $I < 19$ and
the Hubble Deep Field (HDF) counts are systematically high for $I < 21$. 
However, the errors for those surveys are large in those intervals
owing to their small angular coverage and the difference 
is entirely consistent with the observed cosmic fluctuations in counts on
scales of a few arcminutes and probable small differences in isophotal detection
limits. Our agreement with the HDF counts in the range 
$21.7 \lsim I \lsim 24$ is quite important because it suggests that
both the HDF galaxy catalog and ours appear to be counting the same type of 
objects at least in this magnitude range. Given HST's superior resolution, 
this would argue that down to $I = 24$ inclusion of a significant 
population of sub-galactic clumps in the HDF counts 
is not likely (\cite{colley}).

We have normalized the no-evolution models to match our counts precisely
at $I \le 16$ where the models and the data
exhibit the same near-Euclidean slope of 0.55. This is the most appropriate
normalization location since we want to test for 
departures from the no-evolution assumption at cosmological distances. 
Relative to the no-evolution $\Omega = 1$ model at $I = 23$
($0.7 \lsim z_{\rm median} \lsim 1$), 
our counts are consistent with either $\sim 1$ mag of luminosity evolution 
or a factor of $\sim 2$ enhancement in the galaxy density. 
Relative to the $\Omega = 0.02$ model at this same magnitude,
our counts are consistent with either $\sim 0.5$ mag of luminosity evolution 
or a factor of $\sim 1.5$ enhancement in the galaxy density. 
While these models fail to reproduce the very faint Hubble Deep 
Field observations (see
\cite{fb98}), they do reasonably well at brighter magnitudes and thus
provide a plausible comparison in the flux regime covered by our survey.

\section{The Two-point Correlation Function}

\begin{figure}
\epsscale{1.0}
\plotone{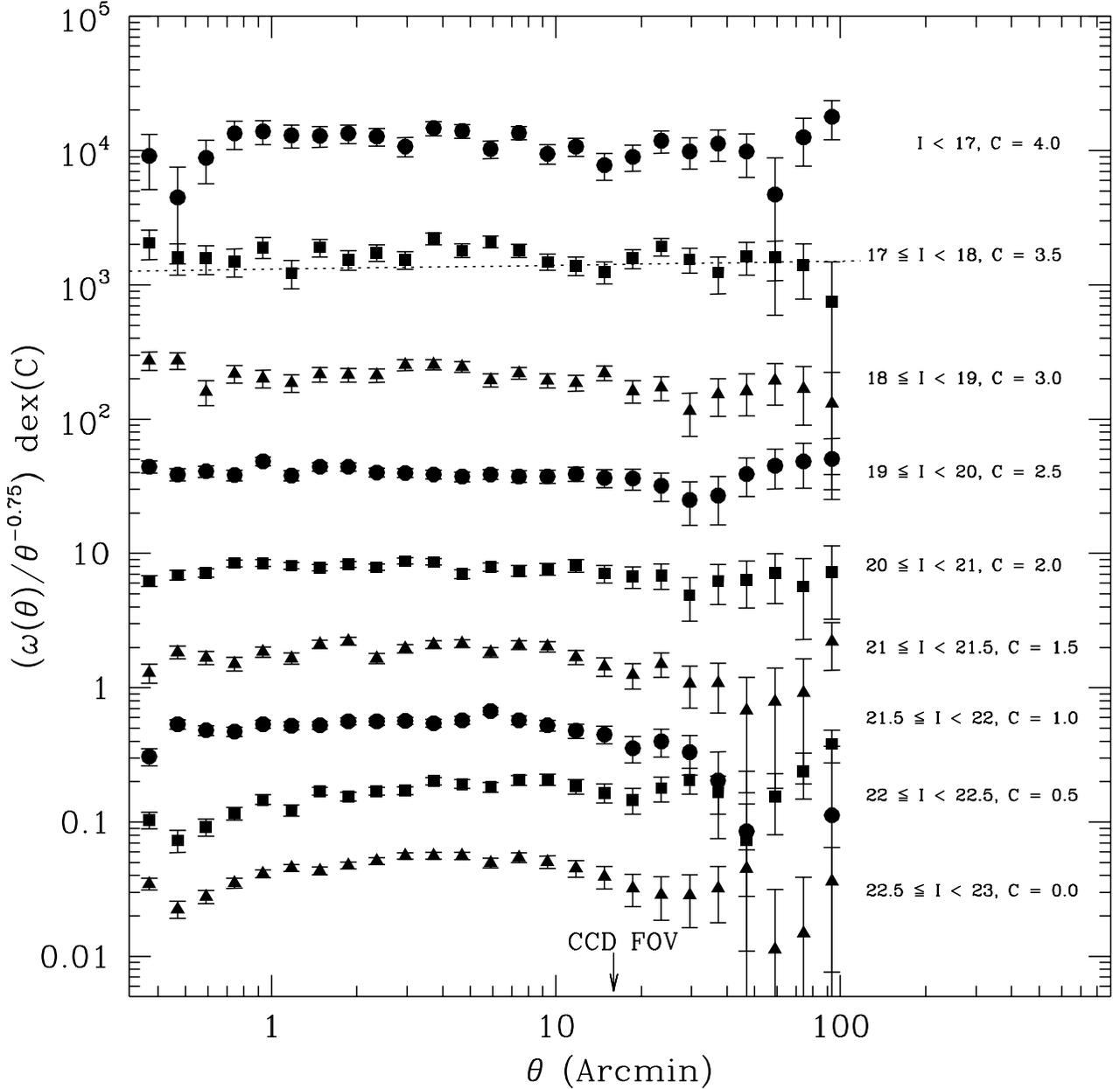}
\caption{The two-point angular correlation functions, normalized
by $\theta^{-0.75}$,
for $17 \le I \le 23$. Results for each magnitude range are vertically
offset by 0.5 dex, as well, for clarity (the
value of each offset is shown explicitly on plot).
The best-fit power law for the
APM $17 < b_J < 20$ galaxy sample is shown as a dashed line (its vertical
offset is the same as that for the $17 \le I < 18$ sample). An arrow
denotes the CCD field of view.}
\label{wtheta}
\end{figure}

We use the \cite{ls93} estimator, $\omega(\theta)=(DD-2DR+RR)/RR$, to 
compute the two-point angular correlation function. Here DD is 
the observed number
of galaxy pairs with separation between $\theta$ and $\theta + \delta\theta$,
RR is the number of such pairs in a randomly distributed sample with
identical boundaries and galaxy surface density, and DR is the number of
such pairs in a cross-correlation between the observed catalog and a random 
realization.  This is the minimum variance estimator and
fully corrects for mask and edge effects up to scales approaching
the survey size. 
Our results for 9 independent magnitude slices, each based on 100 random
realizations,
are shown in Figure~\ref{wtheta} after the application of the
integral constraint (IC) and stellar contamination corrections. 
The IC correction attempts to offset the
artificial reduction in power on large scales that occurs when
the mean galaxy density and $\omega(\theta)$ are determined from 
the same survey (\cite{gp77}). The small reduction in amplitude introduced
by misclassified stars is corrected by multiplying the IC-corrected 
$\omega(\theta)$ values by the factor $N_{Obj}^2/(N_{Obj}-N_{Star})^2$ 
where $N_{Obj}$ is the total number of objects in a given magnitude 
bin and $N_{Star}$ is an estimate of the number of misclassified stars
in the same bin. Since galaxies significantly outnumber stars
when $I > 21.5$, stellar contamination effects are most significant for
brighter magnitudes. However, that is precisely where the PSF-based classifier
works extremely well. Hence we take the number of misclassified stars
to be approximately equal to the square root of the number of stars in the bin. 
We have confirmed that the results from Landy-Szalay estimator 
are in excellent agreement with those from alternative estimators for 
$\omega(\theta)$ such as the counts-in-cells method or the ensemble estimator. 
In each magnitude slice, we only include data from those CCD images in the 
survey that are complete to at least 0.5 mag fainter than the upper mag 
limit of the slice. This assures that large-scale artificial power 
from frame-to-frame depth variations is eliminated. The depth of each frame 
is determined by identifying the location of the peak object counts 
(stars plus galaxies) and then subtracting 0.25 mag.

We find excellent agreement between our results for $17 \le I < 18.25$
and the APM $\omega(\theta)$ determination
(\cite{Maddox}) derived from galaxies with $17 \le b_J \le 20$ (dashed line).
Table 2 presents our correlation function data (multiplied by 100 for
compactness) in each magnitude bin. The data in this table have the
IC and stellar contamination corrections applied.
Table 3 presents the median $I$ magnitudes, number of galaxies each subsample,
best fit power law slopes, the corrected 
amplitudes at 0.5, 1, and 3 arcminutes (interpolated from results in Table 2), 
and the values of the IC and stellar contamination corrections.
The slope of the correlation function is determined from the best
fit to the power law representation
$\omega(\theta) = A_{\omega}\theta^{\delta}$.
The formal error in $\omega(\theta)$ is 
$\sigma_{\omega} = \sqrt{(1 + \omega(\theta))/{\rm RR}}$. This expression
produces error values which are comparable to those estimated by the 
statistical bootstrap method.
The errorbars in Figure~\ref{wtheta} (and Table 2) include an often ignored
additional error term, added in quadrature to the formal error, 
arising from uncertainty in the value of the IC.
The IC uncertainty arises because one typically assumes a power-law form
for $\omega(\theta)$ a priori and thus, on large scales, systematic errors 
can be introduced if this is not an accurate representation of 
the true correlation function. We estimate this
uncertainty by measuring the change in the IC as the assumed
power law slope is varied by $\pm0.1$. 
Strictly, these two error components are not completely independent
(and thus a quadrature addition is not completely proper). However,
the quadrature addition provides a reasonable approximation and,
for the present survey, the IC uncertainty is only important ({\it i.e.},
becomes comparable to the formal error above) on scales
$\theta \gsim 25$ arcminutes.

\subsection{Dependence of the Amplitude on Magnitude}

\begin{figure}
\epsscale{1.0}
\plotone{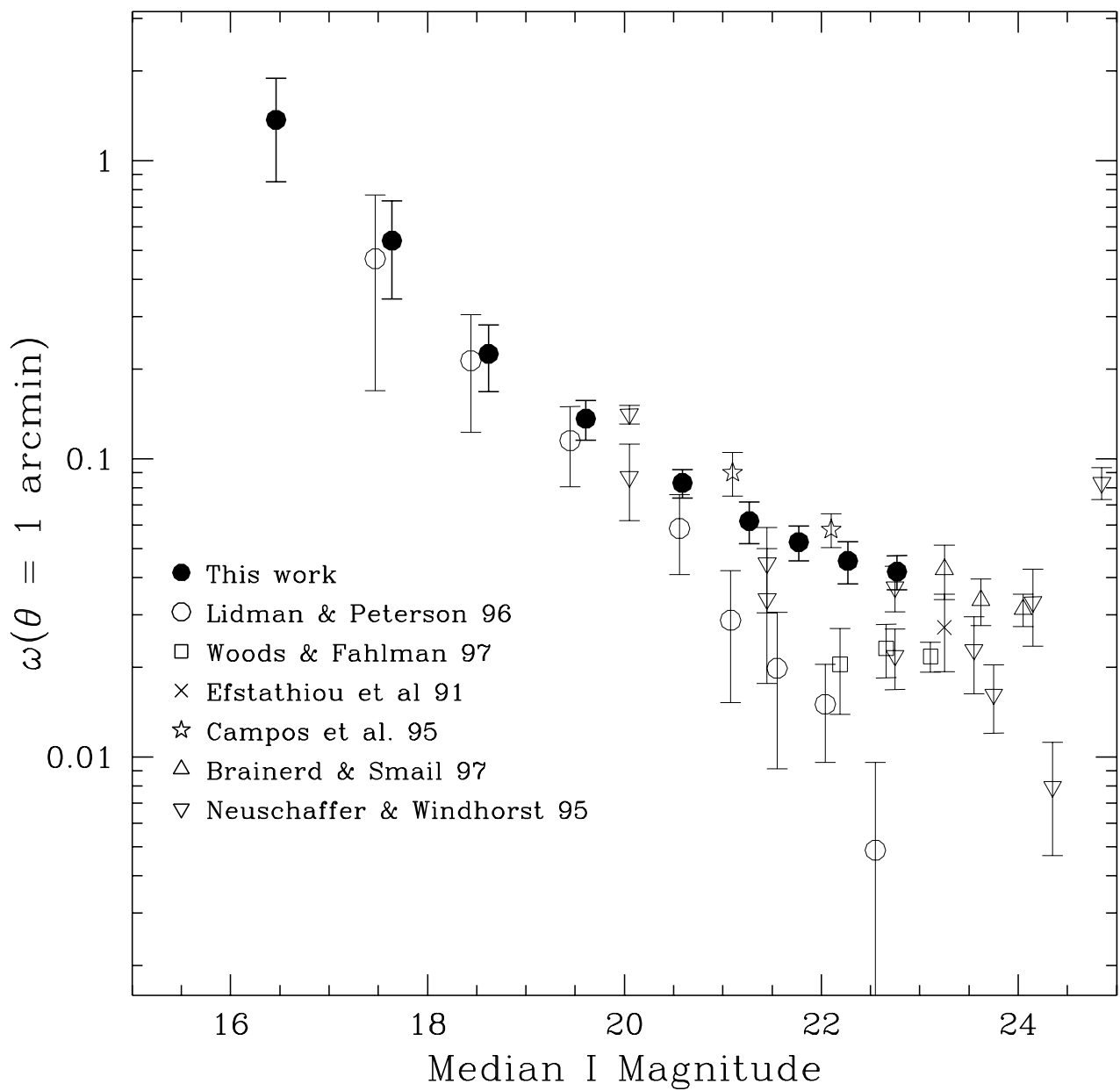}
\caption{Amplitude of $\omega(\theta)$ at $1'$
vs. median $I$ magnitude. Our results are the filled circles. Other
recent measurements are also shown.}
\label{wone}
\end{figure}

Figure~\ref{wone} shows the dependence of the amplitude of $\omega(\theta)$ 
at $1'$ scale on the median $I$ magnitude of the subsample. 
Results from \cite{ef91}, \cite{campos}, \cite{nwind}, \cite{lp96}, \cite{wf97},
and \cite{bs97} are shown for comparison. 
We find a smooth decline in amplitude with median $I$ magnitude
that is consistent with the results from \cite{campos} and
\cite{bs97}.  
For $I \lsim 21$, our results are also consistent with those of 
\cite{lp96}, although their results are systematically lower in
amplitude by 20\% to 45\%. At $I > 22$, we differ at more than the $3\sigma$
level from the results of \cite{lp96} and \cite{wf97}. 

Our results are substantially more reliable than
previous estimates over the range $I < 24$ because our large survey area
overcomes the most significant problem faced by smaller 
surveys: the cosmic variance in galaxy surface density and clustering.
For small surveys, these effects result in large fluctuations in
the correlation amplitude and in a large value of the IC correction on 
arcminute scales. At large scales both of these effects depend on the
average of the correlation function over the survey area.
The ratio of variances expected between two different surveys
is at least $(\theta_1 / \theta_2)^{\delta/2}$,
where $\theta_1$ and $\theta_2$ are the effective angular scales  
of the two surveys and $\delta$ is the slope of the angular
correlation function. While cosmic variance is boosted at
small scales by the higher order correlations neglected
in this simplified formula, the formula  should be a reasonable
accurate approximation for the ratio of finite volume errors.
Thus, relative to many of the previous surveys,
the cosmic variance on arcminute scales in our survey is suppressed 
by more than a factor of 3.
Moreover, the continuity and simple geometry of our survey together
with the large number of galaxies employed ensure that 
edge and discreteness effects are negligible over
a large dynamic range of scales.
The advantage of this large, contiguous area surveys is, therefore,
even better than the above estimate. 

\begin{figure}
\epsscale{1.0}
\plotone{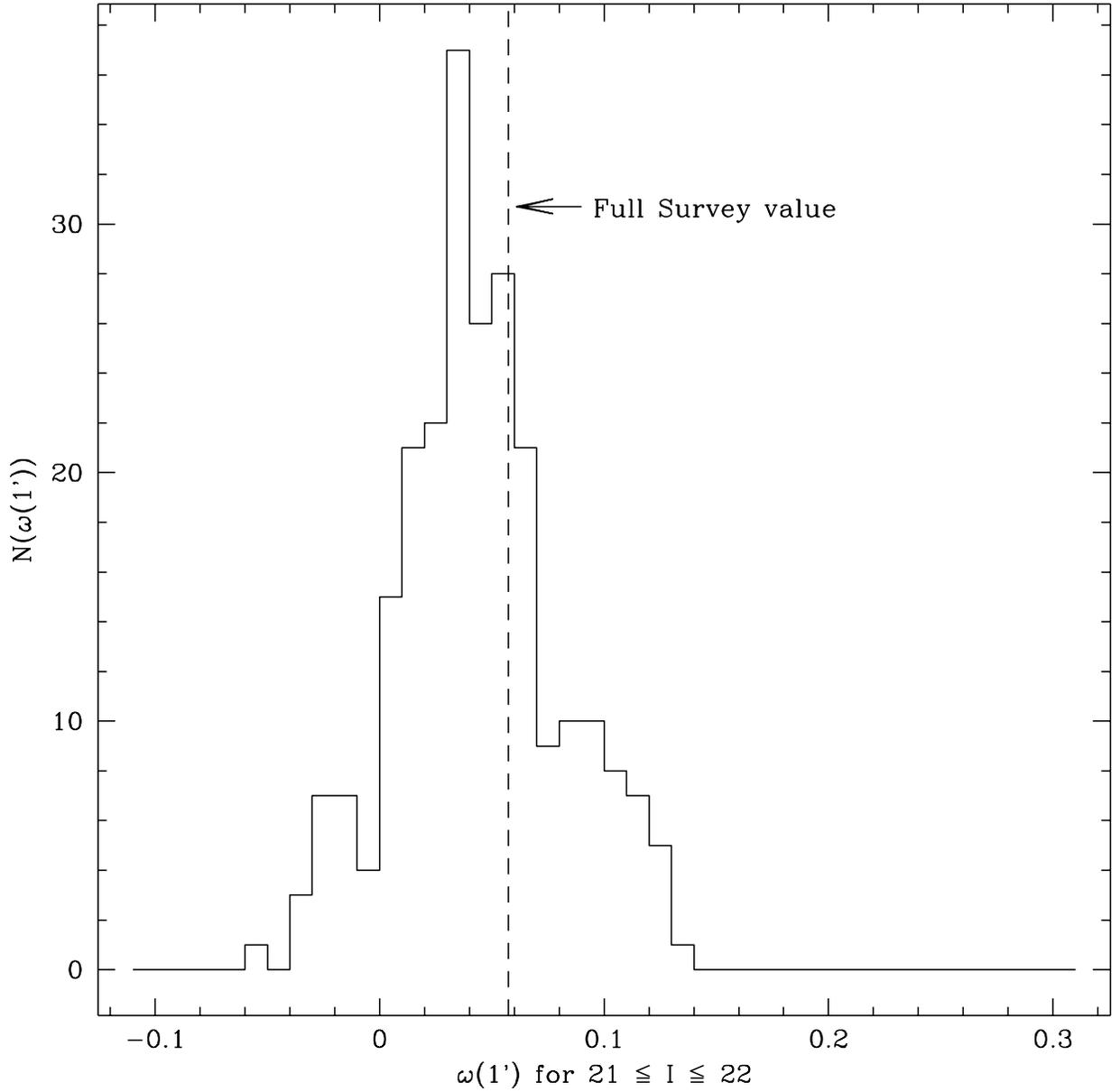}
\caption{The histogram of $\omega(1')$ in the range $21 \le I \le 22$
for 250 independent $16' \times 16'$ CCD fields. The mean value
is 0.045 but the variance is substantial owing to significant cosmic
scatter on this scale at these magnitudes. The $\sim20$\% offset between the
mean $\omega(1')$ for the independent CCD fields and that for
full survey is entirely consistent with the uncertainties in the
small field IC correction.}
\label{wfluc}
\end{figure}

In addition to the above effects, which primarily depend on the 
variance only, the non-Gaussian error distribution for $\omega(\theta)$
is likely to be skewed positively (\cite{sc96}) as well. This can result in
an {\em apparent} low bias in the correlation amplitude for smaller
surveys since there are many areas of sky with
values somewhat lower than the mean, which are not balanced
by the relatively few areas with values
much higher than the mean.
We use our survey data to directly measure the true variance
caused by surface density and clustering properties 
in $\omega(1')$ in many independent but small contiguous areas,
and  find that variance is significant  on 1 arcminute scales. 
We chose a cell size of $16' \times 16'$ --- corresponding to the field
of view of a single KPNO 4m CCD exposure (which is still significantly larger
than the independent fields of previous surveys).
We analyze 250 such fields and computed 
$\omega(\theta)$ for $21 \le I \le 22$,
a magnitude range where we begin to see growing discrepancy between
different surveys. 
The histogram of $\omega(1')$ values is shown in Figure~\ref{wfluc}. 
We find that the scatter in $\omega(1')$
is comparable to its mean value ($\sim0.045$).
Extreme values can reach $3\times$ the mean. 
Thus, in order to measure
$\omega(1')$ with a S/N of 10 one requires at least 100 such
independent fields even at only moderately faint magnitudes.
There is also a 20\% offset between the full survey value and the mean of 
the 250 independent fields, but this difference is entirely consistent 
with uncertainties in the IC correction for the individual CCD frame results.

The IC correction, like cosmic variance, is determined by the
effective survey area, and to lesser extent, by the survey
geometry. The largest contiguous area used
in $\omega(\theta)$ measurements at $I \ge 22$ in the other
works cited is 0.25 deg$^2$ (compared with 14.7 deg$^2$ here).
The IC corrections at $1'$ in our survey are
negligible ($\lsim 0.05\omega(\theta)$) whereas they can
be a significant fraction ($\sim 50 - 100$\%) of the amplitude of
$\omega(\theta)$ in some of the previous surveys ({\it e.g.}, \cite{ef91}).
Hence, while the agreement between \cite{bs97} and our results 
is excellent, their two
survey fields only cover about 0.02 deg$^2$ and, therefore,
additional imaging over larger areas at fainter limits ($I > 24$) is
still desired. The \cite{campos} results are based on a survey
of 64 $11' \times 11'$ non-contiguous 
fields (2.2 deg$^2$ total area) distributed
over an $8^\circ \times 10^\circ$ region.

\subsubsection{Models}

The dependence of the angular correlation function amplitude
on magnitude can be modeled
using the equation (\cite{limber}, \cite{peebles80})
\begin{eqnarray}
\omega(\theta) = \sqrt{\pi} {{\Gamma[(\gamma-1)/2]} \over {\Gamma(\gamma/2)}}
{A(\epsilon) \over {\theta^{\gamma-1}}} r_{o}^{\gamma} 
\end{eqnarray}
that relates $\omega(\theta)$ to the spatial correlation function, 
$\xi(r) = (r/r_o)^{-\gamma}$. A power-law spatial correlation function
implies that it's two dimensional counterpart, $\omega(\theta)$, will
also take on a power-law form. The function $A(\epsilon)$ depends on the
the galaxy redshift distribution, $N(z)$, and on the
evolution of $\xi(r)$. If this evolution is parameterized as
({\it e.g.}, \cite{ef91} and \cite{wf97})
\begin{eqnarray}
\xi(r,z) = ({r \over r_o})^{-\gamma} (1 + z)^{-(3+\epsilon)}
\end{eqnarray}
then $A(\epsilon)$ is
\begin{eqnarray}
A(\epsilon) = \int^{\infty}_{0}D(z)^{1-\gamma}g(z)^{-1}(1+z)^{-(3+\epsilon)}
\left( {dN(z) \over dz} \right)^2 dz \left[\int^{\infty}_{0} 
\left( {dN(z) \over dz} \right) dz \right]^{-2}
\end{eqnarray}
where $D(z)$, the angular diameter distance, and $g(z)^{-1}$ are
\begin{eqnarray}
D(z) = {\rm{c \over H_o}}\left({{q_oz + (q_o-1)(\sqrt{1+2q_oz} - 1)}\over{q_o^2(1+z)^2}}\right)\\ 
g(z)^{-1} = {\rm{H_o \over c}}\left( (1+z)^2\sqrt{1+ 2q_o z}\right)
\end{eqnarray}
In this parameterization, two special cases are noteworthy.
Clustering that remains constant in co-moving coordinates yields
$\epsilon = \gamma - 3 \approx -1.2$. Clustering that
remains fixed with respect to physical coordinates yields
$\epsilon = 0$.

\begin{figure}
\epsscale{1.0}
\plotone{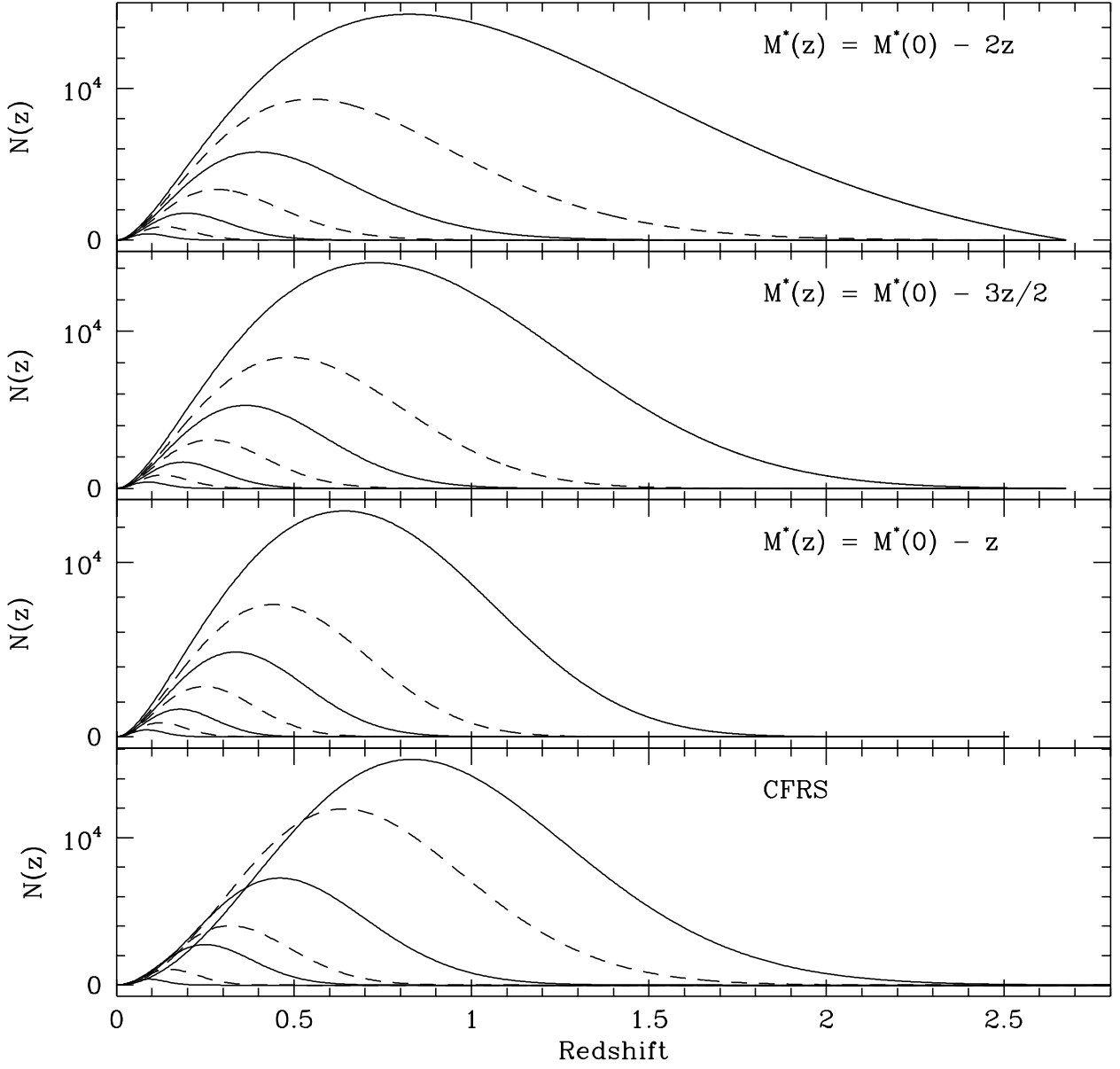}
\caption{The differential redshift distributions for the CFRS model and
the evolving luminosity function (EvLF) models for the magnitude intervals
$16 \le I < 17$,
$17 \le I < 18$,
$18 \le I < 19$,
$19 \le I < 20$,
$20 \le I < 21$,
$21 \le I < 22$, and
$22.5 \le I < 23.5$.
The EvLF models assume a Schechter luminosity function with
$\alpha = -1.1$, $M_{I}^{*}(z) = -20.9 + 5{\rm log}h - \beta z$,
and $\beta = 1,1.5,2$.}
\label{nz}
\end{figure}

We use several different redshift distributions in our fitting procedure
(all shown in Figure~\ref{nz}).
The first is based on fits of the empirical function
$dN(z)/dz \propto z^2 {\rm exp}(-(z/z_o)^2)$ to the CFRS data and its
extrapolation to fainter magnitudes (\cite{CFRS}; hereafter CFRS
model). This parameterization provides a very good representation 
to the observed data (\cite{bs97}) and allows one to use a smoothly varying 
function in the Limber equation, thus removing shot noise effects
present in the actual data. The median redshifts
for the CFRS-based $N(z)$ model are 0.095, 0.145, 0.260, 0.420,
0.540, 0.615, 0.690, 0.755, 0.810, and 0.865 for $I_{median}$ 
mags of 16.5, 17.5, 18.5, 19.5, 20.5, 21.25, 21.75, 22.25, 22.75,
and 23.25, respectively. The remaining $N(z)$ models are generated
by integrating an evolving Schechter luminosity function
in successive redshift shells (hereafter EvLF models) and
within magnitude limits that correspond to our chosen subsamples. 
The evolution is assumed to be a pure luminosity evolution
as represented by the Schechter LF parameters $\alpha = -1.1$, 
$M_{I}^{*}(z) = -20.9 + 5{\rm log}h - \beta z$,
and $\beta = 1,1.5,2$. The $\beta$ values were chosen
to approximately bracket the CFRS model. 
All the models contain a significant fraction of high-$z$ galaxies
by $I = 23.5$: $\sim 30-65$\% with $z > 1$ and $\sim 3-30$\% with $z > 1.5$
for the EvLF models; 44\% with $z > 1$ and 11\% with $z > 1.5$
for the CFRS model.

\begin{figure}
\epsscale{1.0}
\plotone{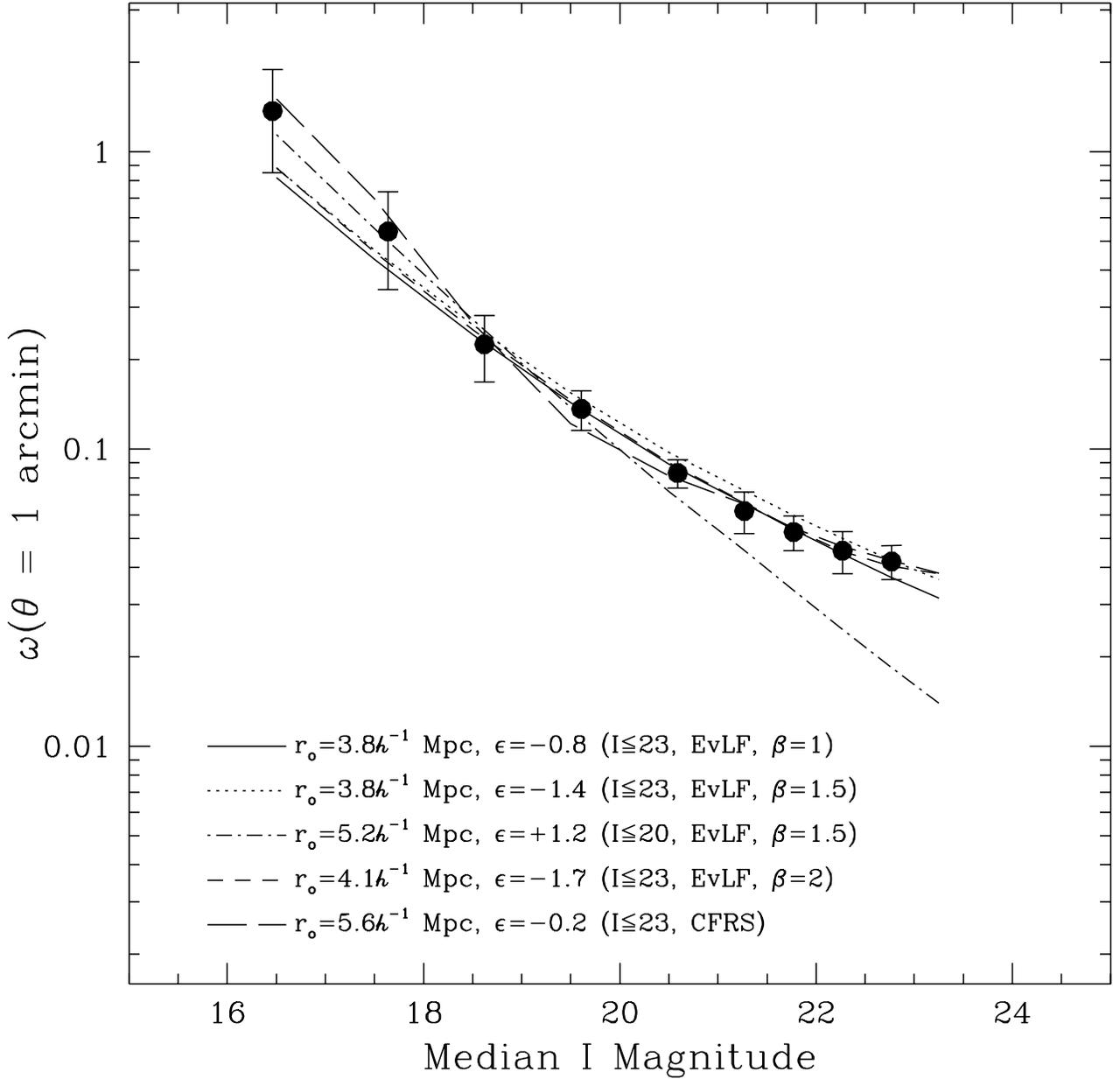}
\caption{Best fits to our $\omega(1')$ vs $I$ mag data
using a Limber deprojection and various plausible redshift distributions.
In the case of the EvLF $\beta = 1.5$ $N(z)$ model, we show fits to all
the data and to just data with $I \le 20$. Fits to data with $I \ge 20$
give similar results as those done using all data.}
\label{wonefits}
\end{figure}

\begin{figure}[ht]
\includegraphics{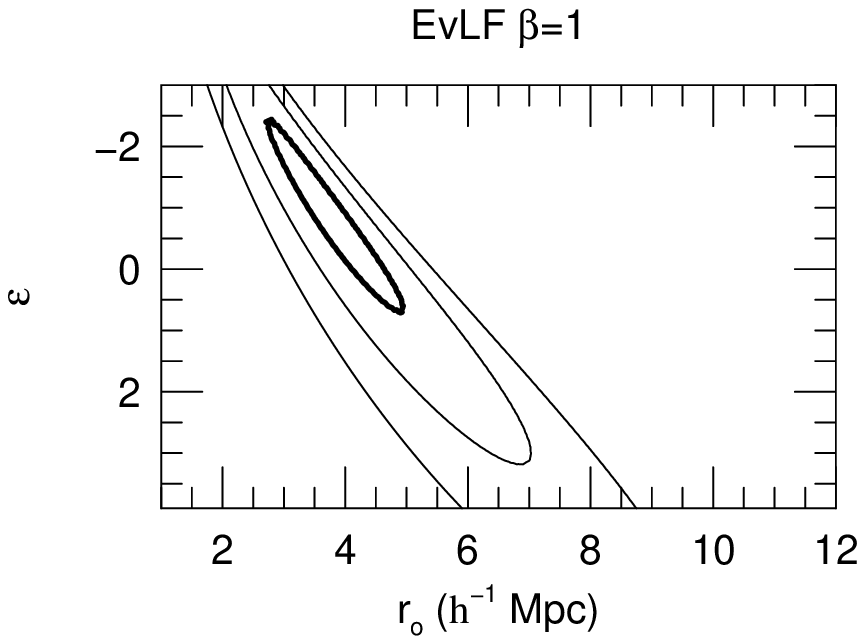}
\includegraphics{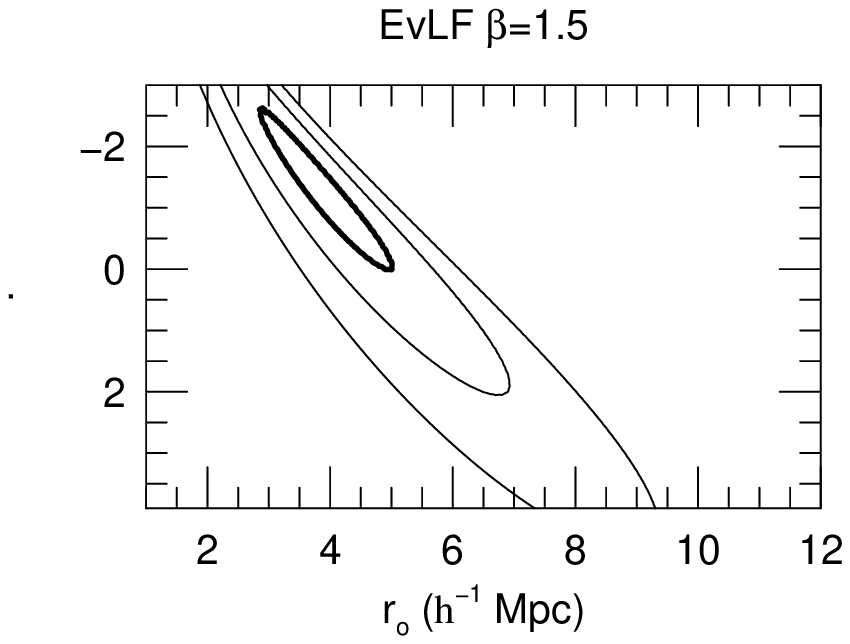}
\includegraphics{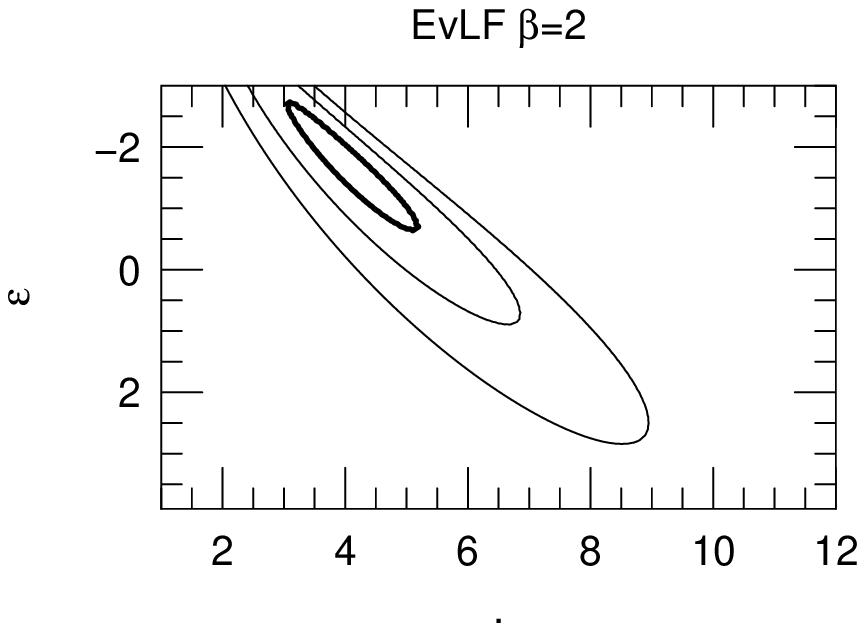}
\includegraphics{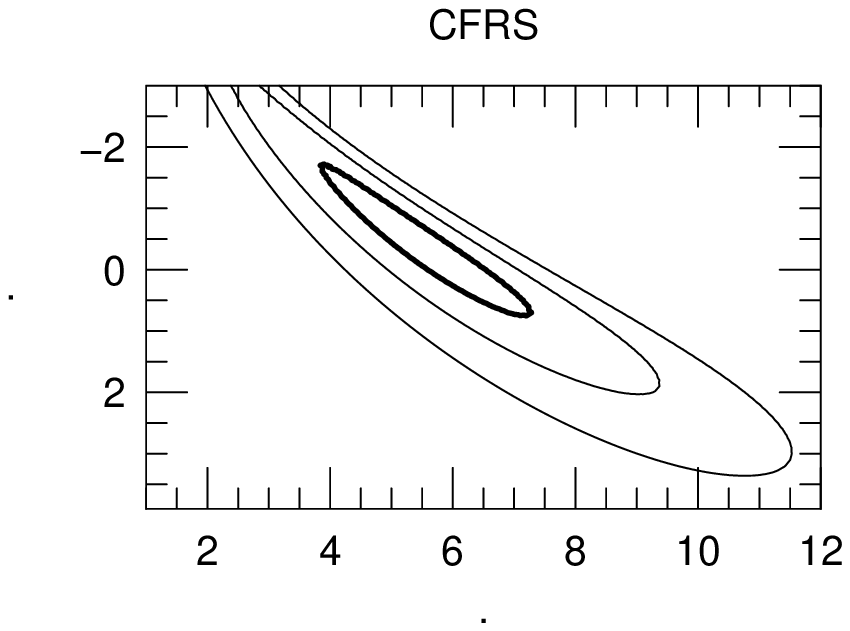}
\vskip 7.0in
\caption{The reduced $\chi^2$ contours for the fits to
equation 1 using
the EvLF ($\beta = 1, 1.5, 2$) and CFRS redshift distributions.
Results shown here are for correlation amplitudes at 1 arcminute
and $I_{\rm median} \le 23$.
The $1\sigma$ (heavy line), $2\sigma$, and
$3\sigma$ contours are shown.}
\label{chisqr}
\end{figure}

For each $N(z)$ model, we find the best-fit to our $\omega(1')$ vs $I$ 
magnitude data by identifying the combination
of $r_o$ and $\epsilon$ that minimize the $\chi^2$ statistic.
Table 4 and Figures~\ref{wonefits} and \ref{chisqr}
summarize the results of the fitting procedure. The best
fit in the range $16 < I < 23$ using the CFRS $N(z)$ model 
gives $r_o = 5.6h^{-1}$ Mpc and $\epsilon = -0.20$. The best
fits for the EvLF $N(z)$ models are $(r_o,\epsilon) = $
$(3.8h^{-1}\ {\rm Mpc}, -0.80)$, 
$(3.8h^{-1}\ {\rm Mpc}, -1.40)$,  and
$(4.1h^{-1}\ {\rm Mpc}, -1.70)$ for $\beta = $1, 1.5, and 2, respectively.
If we fix the correlation length to be $5.5\pm1.5h^{-1}$ Mpc, a value that
is typical of that found from local redshift surveys ({\it e.g.},
\cite{delap}, \cite{jing}, \cite{tucker}), the
best fit $\epsilon$ values fall in the range $-0.4 \le \epsilon \le +1.3$.
Restricting the fits to data with $I \le 20$, in general, also
yields more positive $\epsilon$ values and higher correlation lengths that
are consistent with those from the above local redshift surveys (see 
columns 3 and 5 in Table 4). 
Fits to data in subsets between $I = 18$ and $I = 23$
are consistent with the results obtained using all data with $I \le 23$.
In other words, we see little change in the correlation length with
magnitude for $18 \le I \le 23$ for a given redshift distribution.
All fits were done assuming
a slope of $\gamma = 1.8$ and q$_{\rm o} = 0.1$ 
(for q$_{\rm o} = 0.5$ the best fit $r_o$ values are $\sim 7$\% lower).

Figure~\ref{chisqr} shows the reduced $\chi^2$ contours 
for the four $N(z)$ models as functions of $r_o$ and $\epsilon$.
Two things are clear from the $\chi^2$ measurements. First,
there is a significant covariance between $r_o$ and $\epsilon$ -- for a given
redshift distribution, lower correlation lengths are coupled with lower
$\epsilon$ values. 
Second, the best-fit is located within a fairly shallow minimum at least
in the direction of the $r_o - \epsilon$ covariance.   
Thus, although the best fits prefer $\epsilon < 0$, positive values
are included in the $1\sigma$ contours, albeit with
correspondingly higher correlation lengths. We provide an approximate
description of the covariance between $r_o$ and $\epsilon$ by
determining the best-fit line to the data within the $1\sigma$
contour. The fit is determined by computing the mean between
the fits $r_o = f(\epsilon)$ and $\epsilon = f(r_o)$. 
The typical $1\sigma$
uncertainty in $r_o$ at a given $\epsilon$ is $\pm0.15h^{-1}$~Mpc 
and $\pm0.23h^{-1}$~Mpc 
for the EvLF and CFRS models, respectively (for the full $I \le 23$
dataset). The typical $1\sigma$ 
uncertainty in $\epsilon$ at a given $r_o$ is $\pm0.18$ (for all models).
These fit parameters are given in Table 4.

\subsection{Dependence of $\epsilon$ on Angular Scale}

\begin{figure}
\epsscale{1.0}
\plotone{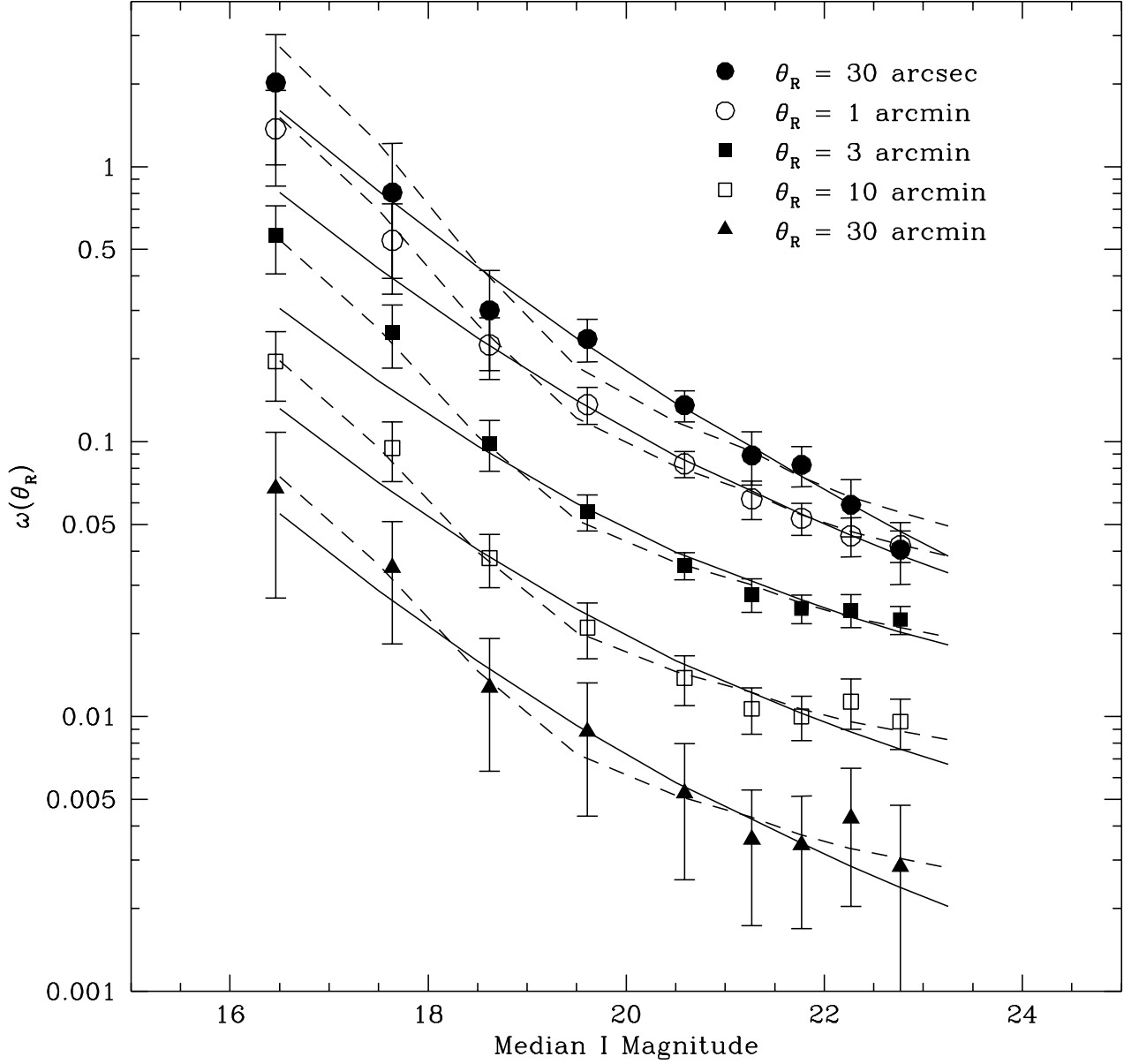}
\caption{The amplitude of $\omega(\theta_R)$ as a function of median
$I$ magnitude for $\theta_R = 0.5', 1', 3', 10'$, and $30'$. Data
have been corrected for integral constraint and stellar contamination.
Best-fit models using the EvLF $\beta = 1.5$ and CFRS redshift distributions
are shown as solid and dashed lines, respectively.}
\label{allscales}
\end{figure}

\begin{figure}
\epsscale{1.0}
\plotone{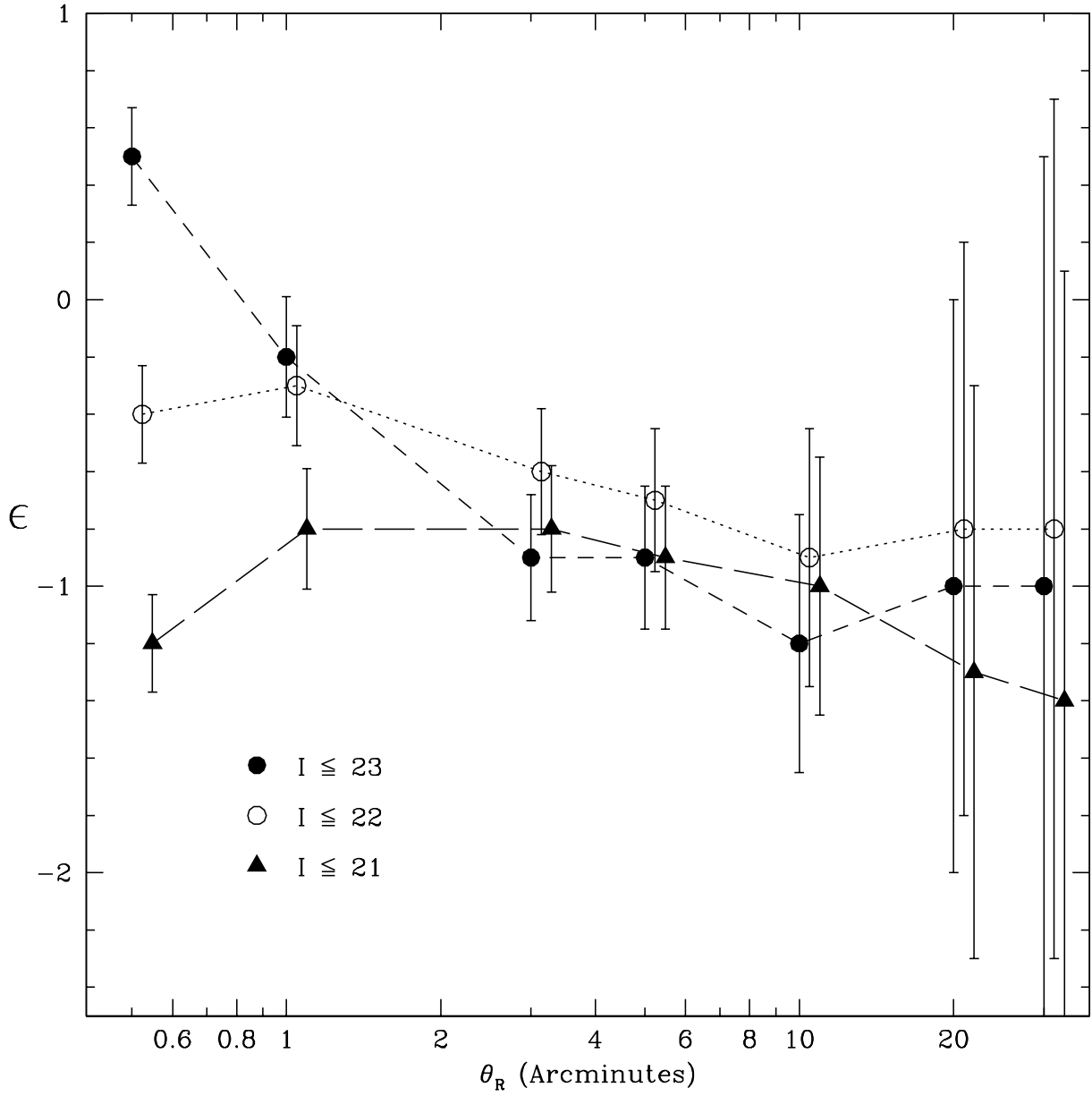}
\caption{The best-fit $\epsilon$ value as a function of angular scale
for data with $I \le 21$, $I \le 22$, and $I \le 23$.
The results shown are for the CFRS redshift distribution model. Results
for the EvLF models are similar in shape but are offset vertically
relative to the CFRS fits.  }
\label{eps_allscales}
\end{figure}

Our large survey allows us to perform model
fits at scales other than $1'$. We perform a similar analyses for
$\theta = 0.5', 3', 10'$, and $30'$. The results are shown in
Figure~\ref{allscales}. Figure~\ref{eps_allscales} shows the best fit
$\epsilon$ value as a function of angular scale obtained when the CFRS redshift
model is used. The results for the EvLF redshift models are similar shape 
although there is a zeropoint shift in the $\epsilon$ values. For $I > 22$,
there is a trend towards larger $\epsilon$ values as the angular scale decreases, 
perhaps suggesting that the clustering on the smallest scales evolves more rapidly.
The uncertainties in $\epsilon$ on scales greater than 10 arcminutes
are substantial, however, and the trend
is not seen at $I \le 21$. We show in the following section that there
is a significant flattening of the slope of the correlation function
when $I > 22$ as well. These two trends are not easily identified with
any known observational or instrumental effects (see discussion below) 
but clearly a deeper, wide area survey is needed to confirm these results.
 
\subsection{Dependence of the Slope on Magnitude}

\begin{figure}
\epsscale{1.0}
\plotone{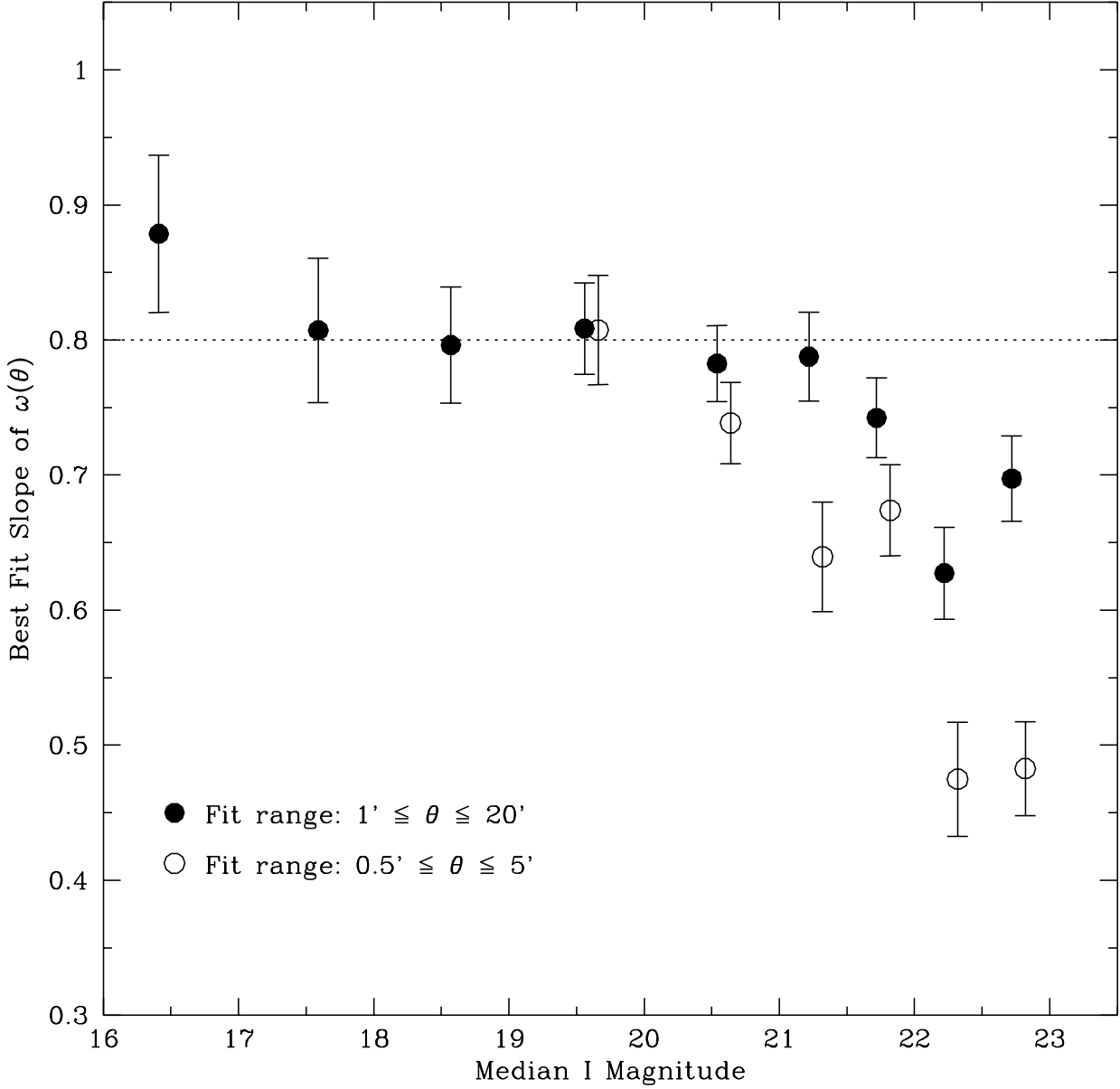}
\caption{The best fit power law slope of the two-point correlation
function as a function of $I$ magnitude. The results for two different
angular fitting limits are shown. The dashed line shows the mean slope
derived for $I \le 22$.}
\label{acfslope}
\end{figure}

We find no significant dependence of the correlation function slope on magnitude
for $I \le 22$ as shown in Figure~\ref{acfslope}. Over this magnitude range
and for $1' \le \theta \le 20'$ the best fit slope is $\delta = -0.80 \pm 0.02$.
For our two faintest bins ($I > 22$), a flattening of the slope on angular 
scales $\theta \lsim 5'$ is seen
with $\delta = -0.48 \pm 0.04$ for $22 \le I < 23$
(see Figure~\ref{wtheta} and Table 3 as well). 
The trend is not explained by seeing
or deblending effects as it is also seen in the subset of the data with
the best resolution (FWHM $\le 1.25''$). It could, in principle, be introduced
by poorly calibrated frame to frame photometric zeropoint variations. However,
our observing strategy enables us to measure and correct for these 
variations quite well (see \S2) and any residual 
variations are not sufficient to introduce a flattening of this magnitude. 
Furthermore, the $22 \le I < 22.5$ bin is still 1 magnitude brighter than the
completeness limit. Experiments with simulated CCD data show that our object 
detection software is not dependent on the clustering properties of the objects. 
As an additional check, however, we compute $\omega(\theta)$ 
independently for the 4 quadrants of the survey, for the single run
with the best atmospheric transparency, and for each individual CCD
frame. We see the flattening in these subsamples as well.
One would not expect an effect introduced by zeropoint miscalibration
to survive all these experiments. This dependence has been seen in two other
surveys at approximately the same magnitude and over the same angular
scales as seen here. \cite{nwind} report slopes of $\delta=-0.5$ by $g=25$
based on two independent fields, each one covering 0.25 contiguous deg$^2$. 
\cite{campos} find best fit slopes of $\delta=-0.6$ and $\delta=-0.55$
for their $R \le 22$ and $R \le 23$ samples, respectively. 
There is a physical model that can be invoked to
explain the flattening; we will summarize it in the next section.
While it is intriguing that our results are roughly consistent with those
seen in these two independent surveys done at different wavelengths,
we remain cautious for now since the effect is only seen 
in our faintest bins.
Other large, contiguous surveys now underway (\cite{noao}, \cite{btc50})
should provide an important check on this result.
We note that the constraints on $r_o$ and $\epsilon$ discussed above do
not change if we exclude the two faintest points from the analysis. 

\section{Discussion}

Our constraints on $\omega(\theta)$ reveal a dependence of its amplitude
on median $I$ magnitude that appears to fit the model
of the redshift evolution of $\xi(r)$ parameterized
as $\xi(r,z) = ({r \over r_o})^{-\gamma} (1 + z)^{-(3+\epsilon)}$.
In fact, this model provides an acceptable representation of the data
over the range $16 < I < 25$  if the results
of \cite{bs97} are also included in the fits. The decline in the amplitude of 
the correlation function (over a wide range of scales) with $I$ magnitude
is about 3 times steeper over the range $16 \lsim I \lsim 19$ than
over the range $20 \lsim I \lsim 23$. 
The $N(z)$ distributions that yield good fits to magnitude-$\omega(\theta)$
relation typically have about 40\% of the galaxies at
$z > 1$ by $I = 23$, consistent with the conclusions of
\cite{bs97}. A shallower decline in the amplitude of
$\omega(\theta)$ with magnitude for $I \gsim 21$
can be induced by the reduction of the proper volume element at higher $z$.
\cite{bs97} use this argument to support the assertion that a significant
fraction of faint galaxies are at $z \gsim 1$. However,
a high-$z$ cutoff can also induce similar behavior since in this
case the effective depth and redshift distribution are not changing
dramatically as one goes to fainter limits. If such a cutoff were
introduced, as noted by \cite{lp96}, by a selection bias
{\it against} the detection of high-$z$ galaxies
a shallow decline in correlation length with magnitude
would result. However, a strong redshift cutoff would also effect the 
number counts and the corresponding turnover in the counts is not seen 
(see also \cite{wf97}). Given the excellent agreement we find with the 
HDF number counts, we conclude that a selection bias of this nature is
not a significant problem for our sample.

Taken at face value,
our best fits to the above model suggest the evolution of the spatial 
correlation function is relatively mild over the depth of the survey. 
When we allow $r_o$ to be a free parameter in the fit, we obtain exponents in 
the range $-1.7 \lsim \epsilon \lsim -0.2$ with $3.8 \lsim r_o \lsim 5.6h^{-1}$ 
Mpc depending on the assumed redshift distribution.
Growth in the clustering of galaxies (in proper coordinates) requires
$\epsilon > 0$.
The 95\% confidence limits are substantial, nevertheless, given the strong 
covariance between $r_o$ and $\epsilon$. As Figure~\ref{chisqr} 
demonstrates, our data are also consistent with $\epsilon > 0$ providing 
$4.5 \lsim r_o \lsim 7h^{-1}$ Mpc. 
If $\epsilon \gsim 1$ then values of $r_o \lsim 4h^{-1}$ Mpc are strongly
rejected. As already noted, redshift surveys of local galaxies find 
$r_o \sim 5.5\pm1.5h^{-1}$ Mpc.

The interpretation of the goodness of the fit to the above model, however,
must take into account a number of additional selection effects that may
mask more substantial evolution of the correlation function.
It is known that low-$z$ galaxy clustering depends on both the morphology
and luminosity of the objects being studied
({\it e.g.}, \cite{dg76}, \cite{moore94}, \cite{love95},
\cite{guzzo97}, \cite{vl97}). 
The sense of these trends is that elliptical galaxies and more luminous
galaxies tend to have larger correlation lengths than spiral galaxies and
less luminous galaxies, respectively. 
We also know that the mean absolute luminosity
of the objects in a flux-limited sample will increase with redshift.
The relative insensitivity
of our derived $r_o$ values to $I$ magnitude for galaxies with 
$I \ge 17$ could therefore be, in part, the result of two competing effects. 
As one goes fainter the mean redshift of the galaxies in
the survey tends to increase. 
If galaxies at $0.5 \lsim z \lsim 1$ exhibit a similar luminosity dependent
clustering as local galaxies and/or $I-$band selection enhances the elliptical
galaxy fraction at these redshifts then it is possible that a decrease in the 
spatial correlation length with magnitude 
could be partially offset by the tendency for more luminous and/or
early type galaxies to be more strongly clustered. 
This could also explain the observed preference for negative $\epsilon$
values (which suggests only modest clustering evolution).
The precise evolution of $\epsilon$ and $r_o$ with redshift
is intimately connected to the density pertubation power
spectrum and the galaxy merger rate (\cite{mosc98}). The complexity
of the dependence makes a unique theoretical interpretation the
observational constraints problematic.
None the less, the constraints are now solidly determined on degree-scales and
less down to $I = 23$ from the present survey.

More stringent constraints on clustering evolution will require 
the addition of multiple passbands or redshift data.
\cite{ef91} report a marginal increase
in the clustering amplitude as one selects redder passbands and posit
that normal galaxies dominate the composition of fainter $I-$band
selected galaxy samples as opposed to the weakly clustered
faint galaxies, that appear to dominate $U$ and $B$ selected surveys.
\cite{lp96} also find a marginally significant
color dependence, showing that red ($V-I > 1.5$)
galaxies have a larger clustering amplitude than blue galaxies. 
\cite{lefev96} find no color dependence to the clustering properties
of galaxies at $z > 0.5$ from an analysis of the CFRS data.  
They have also measured the spatial correlation function directly
from their redshift data. They find a correlation length of 
$r_o = 1.57\pm0.09h^{-1}$ Mpc  at $z \sim 0.5$ which, if they are
sampling a mix of galaxies similar to those in local redshift surveys,
implies $0 < \epsilon < 2$. Our results are consistent with this 
$\epsilon$ range. From equation (2), we know 
$r_o(z) = r_o(0) \left( 1 + z \right)^{(-(3+\epsilon)/\gamma)}$.
Using our best-fit results for the CFRS redshift distribution
we, thus, find that $r(z=0.5) \approx 3.0h^{-1}$ Mpc. Results for the
EvLF redshift distribution models yield $r(z=0.5) = 2.7 \pm 0.4h^{-1}$
Mpc. The CFRS galaxies were also 
$I-$band selected so any difference in results between
the CFRS and our angular survey are due either to
projection effects (which tend to wash out clustering)
or the volume sampled
as opposed to intrinsic differences between the galaxy population.

\cite{ajc98} determined the redshift dependence of the
amplitude of $\omega(\theta)$ using photometric redshifts derived
from the HDF survey. For $z > 0.4$, they
find $r_o = 2.37h^{-1}$ Mpc and $\epsilon = -0.4^{+0.37}_{-0.65}$
However, in order to be consistent with low-$z$ surveys, they would
require $\epsilon \sim 2$. They suggest, therefore, that the expression for
the evolution of the spatial correlation function, 
$\xi(r,z) = ({r \over r_o})^{-\gamma} (1 + z)^{-(3+\epsilon)}$, is
not particularly good. Others have made similar claims based on
poor fits to the $\omega(\theta)$ vs. magnitude relation. 
This is in contrast with the excellent fit we obtain
extending over $7$ magnitudes (Figure~\ref{wonefits}). 
We speculate that the discrepancy is caused by cosmic variance
due to the small area used in previous measurements of the
correlation function.
\cite{ajc98} note that the HDF survey subtends only $800h^{-1}$ kpc
at $z = 1$ (q$_o = 0.1$), a fraction of the
galaxy correlation length. The CFRS is larger (five $10' \times 10'$ fields)
but still covers a relatively small volume ($10'$ spans $3.1h^{-1}$
Mpc at $z = 1$).  Hence, these surveys are also subject to the same sorts
of problems associated with cosmic scatter and undersampling of LSS
as small area angular surveys. Indeed, \cite{delap} make a clear 
demonstration of this point at low-$z$: even in their redshift survey
of $\sim 1810$ galaxies ($z \le 0.05$)
filling a volume of $5 \times 10^5h^{-3}$ Mpc$^3$, the density
fluctuations caused by LSS prevent the determination of $r_o$ by better
than a factor of 2. Given that the deep, redshift probes sample roughly
comparable volumes (but with the added complication that they stretch
over a significant fraction of cosmic time), it seems optimistic
to expect reliable constraints to be determined from the existing surveys.
In fact, it is quite remarkable that the values obtained from them
are in agreement with the local correlation length to within a factor of
two. Not until much larger volumes are surveyed will the constraints on
$r_o$ at high redshift be robustly measured. 
If the results from direct measurements of the spatial correlation function
at higher redshifts are ultimately accurate, however, 
and assuming their sample composition is not
significantly different from the low redshift surveys, then
the basic model may be in need of refinement. 

One possible refinement, discussed in some detail already by \cite{nwind},
is to include a redshift-dependent spatial correlation function slope. 
This would be a natural consequence of models in which there is a 
scale dependence to the growth of structure. For example, if galactic 
scale structures grow
significantly faster than structures on scales $\gsim 20h^{-1}$ Mpc the
correlation function slope would steepen with time. This would also 
cause the slope of $\omega(\theta)$ to flatten at fainter magnitudes on
small scales and, depending on the details of the density perturbation power
spectrum and the merger rate, could also result in an increase in $\epsilon$
with redshift on small scales. 
Adopting the parameterization $\gamma(z)=1.8(1+z_{med})^{-C}$ 
proposed by \cite{nwind}, implies $C = 0.35\pm0.10$ which is consistent
with their limit. 
As \cite{nwind} demonstrate, values in the range $C \lsim 0.4$, are consistent
with structure formation models which have significant power 
on co-moving scales of $\lsim 10h^{-1}$ Mpc but disfavor models with little
small scale power such as HDM. However, since the flattening is only seen at
$I \gsim 22$, the above parameterization of the slope redshift dependence
is not particularly good -- there appears to be no real redshift dependence
to the slope until $z_{med} \gsim 0.6$. 

\section{Summary}

We have produced an $I-$band selected catalog of $\sim 710,000$ 
galaxies for studying the evolution of structure out to $z \sim 1$. 
Our survey puts new and stringent limits on the 
magnitude dependence of the two-point angular correlation function.
The survey is the first of a new generation of deep, wide-area 
imaging surveys made feasible by large-format CCDs and mosaic cameras.
The key results presented in this paper are
\begin{enumerate}

\item Galaxy number counts are consistent with modest evolution in 
luminosity ($0.5-1$ mag, depending on $\Omega_o$)
and/or density (factor of $1.5-2$). The counts are consistent with 
the F814W HDF counts in the range $21 < I \le 24$ suggesting
that the HST-based catalog does not contain a substantial number of
sub-galactic components in this flux range.

\item The two-point correlation function agrees remarkably
well with that of the APM in the magnitude range common to both
surveys. At the faint end, our measurements of $\omega(\theta)$
are consistent with those of \cite{bs97}, who have measured $\omega(\theta)$ 
down to $I = 25$. The large contiguous size of our survey minimizes
the effects of cosmic scatter and provides a firm determination
of $\omega(\theta)$ over 7 magnitudes.

\item The amplitude of the two-point angular correlation function
decreases monotonically with increasing magnitude over the range
$16 \le I \le 23$. The decline is well fit by a redshift dependence
of the spatial correlation function parameterized as $(1+z)^{-(3+\epsilon)}$.
Our best fit correlation lengths for galaxies with $I \le 20$ 
are consistent with those from low-$z$ redshift surveys. At fainter
magnitudes we find correlation lengths in the range $3.8 \le r_o \le 5.5h^{-1}$
Mpc depending on the assumed redshift distribution. If $\epsilon \gsim 1$
then we strongly reject $r_o \lsim 4h^{-1}$ Mpc for our $I > 20$ sample. 
However, more negative $\epsilon$ values are preferred at $I > 20$, suggesting
the effective clustering is not evolving as rapidly as linear perturbation
theory predicts. The possible preference
for inclusion of more luminous and/or early type galaxies at fainter magnitudes
in the survey could explain this trend.

\item Our derived correlation length {\it at} $z \approx 0.5$ 
is about a factor of 2 larger than that derived from the CFRS. While
a complete physical interpretation of our results requires
additional color and redshift data, it is also clear that the volumes
sampled by existing deep, redshift probes are inadequate to properly
constrain the two-point spatial correlation function to better than the
above difference. If the galaxy population sampled at $I \gsim 20$ is
similar to that in the CFRS, then our deprojection of $\xi(r)$ may indeed
be more accurate by virtue our reduced sensitivity to cosmic scatter.

\item For $I \le 22$, the mean value of power-law slope of $\omega(\theta)$
is $-0.80 \pm 0.02$ and is independent of 
magnitude. We find a slope of $\delta \approx -0.5$ is a better fit to 
the $I > 22$ data, however. This flatter slope is
consistent with the results of \cite{campos} and \cite{nwind}.
We also detect an increase in $\epsilon$ on scales less than 1 arcminute
for $I > 22$.
We can identify no instrumental or software-related cause for these
effects but remain cautious in interpreting the full signals as real phenomena.
Such trends are, however, consistent with structure formation models in which
small scale power increases more rapidly than power on scales larger than
$\sim 10h^{-1}$ Mpc. 

\end{enumerate}

Special thanks go to the KPNO TAC for their generous support of
this research. We thank John Hoessel for help with the
initial project proposal and Ralph Shuping for providing essential
photometric calibration data. We thank the referee for helpful comments
which improved the clarity of this paper. M.P. acknowledges support from the
STScI's Director's Discretionary Research Fund.
I.S. was supported by DOE and NASA through grant
NAG-5-2788 at Fermilab and by the PPARC rolling grant for 
Extragalactic Astronomy and Cosmology at Durham.

\renewcommand{\baselinestretch}{1}
\pagestyle{empty}

\voffset -0.5in
\hoffset -0.25in
\newpage

\newpage

\newpage

\newpage

\newpage
 
\newpage

\clearpage
 
\newpage
\hoffset -0.25in
\voffset -0.5in

\newpage

\newpage
 
\newpage
\voffset -0.75in
\hoffset -0.90in
 
\begin{deluxetable}{rrrrr|rrrrr}
\tablewidth{0pt}
\tablenum{1}
\tablecaption{Differential $I$-band Galaxy Counts}
\tablehead{
\colhead{$I$ mag} &
\colhead{log(N)\tablenotemark{a}}  &
\colhead{log(N$_{\rm Cor}$)\tablenotemark{a}} &
\colhead{$\sigma_{high}$\tablenotemark{b}}  &
\colhead{$\sigma_{low}$\tablenotemark{b}}  &
\colhead{$I$ mag} &
\colhead{log(N)\tablenotemark{a}}  &
\colhead{log(N$_{\rm Cor}$)\tablenotemark{a}} &
\colhead{$\sigma_{high}$\tablenotemark{b}}  &
\colhead{$\sigma_{low}$\tablenotemark{b}}}
\startdata
13.25 &-0.5671 &-0.5671 & 0.3010 & 0.5671& 18.75 & 2.9433 & 2.9433 & 0.0076 & 0.0077  \nl
13.50 &-0.2661 &-0.2661 & 0.2323 & 0.5333& 19.00 & 3.0641 & 3.0641 & 0.0066 & 0.0067  \nl
13.75 & 0.3360 & 0.3360 & 0.1315 & 0.1895& 19.25 & 3.1613 & 3.1613 & 0.0059 & 0.0060  \nl
14.00 & 0.3872 & 0.3872 & 0.1249 & 0.1761& 19.50 & 3.2696 & 3.2696 & 0.0052 & 0.0053  \nl
14.25 & 0.4329 & 0.4329 & 0.1193 & 0.1651& 19.75 & 3.3633 & 3.3633 & 0.0047 & 0.0047  \nl
14.50 & 0.7551 & 0.7551 & 0.0857 & 0.1069& 20.00 & 3.4547 & 3.4547 & 0.0042 & 0.0043  \nl
14.75 & 0.8801 & 0.8801 & 0.0752 & 0.0910& 20.25 & 3.5482 & 3.5482 & 0.0038 & 0.0038  \nl
15.00 & 1.0240 & 1.0240 & 0.0645 & 0.0758& 20.50 & 3.6360 & 3.6360 & 0.0034 & 0.0035  \nl
15.25 & 1.1142 & 1.1142 & 0.0586 & 0.0677& 20.75 & 3.7218 & 3.7218 & 0.0031 & 0.0031  \nl
15.50 & 1.2590 & 1.2590 & 0.0501 & 0.0566& 21.00 & 3.8007 & 3.8007 & 0.0028 & 0.0029  \nl
15.75 & 1.3967 & 1.3967 & 0.0431 & 0.0478& 21.25 & 3.8904 & 3.8904 & 0.0026 & 0.0026  \nl
16.00 & 1.5665 & 1.5665 & 0.0357 & 0.0389& 21.50 & 3.9690 & 3.9690 & 0.0023 & 0.0023  \nl
16.25 & 1.6582 & 1.6582 & 0.0323 & 0.0349& 21.75 & 4.0486 & 4.0486 & 0.0021 & 0.0021  \nl
16.50 & 1.7870 & 1.7870 & 0.0280 & 0.0299& 22.00 & 4.1196 & 4.1437 & 0.0019 & 0.0019  \nl
16.75 & 2.0046 & 2.0046 & 0.0219 & 0.0231& 22.25 & 4.1825 & 4.2000 & 0.0027 & 0.0027  \nl
17.00 & 2.0985 & 2.0985 & 0.0197 & 0.0207& 22.50 & 4.2579 & 4.2820 & 0.0024 & 0.0024  \nl
17.25 & 2.2175 & 2.2175 & 0.0173 & 0.0180& 22.75 & 4.3222 & 4.3554 & 0.0022 & 0.0022  \nl
17.50 & 2.3327 & 2.3327 & 0.0151 & 0.0157& 23.00 & 4.3873 & 4.4385 & 0.0020 & 0.0020  \nl
17.75 & 2.4875 & 2.4875 & 0.0127 & 0.0131& 23.25 & 4.4549 & 4.5438 & 0.0018 & 0.0018  \nl
18.00 & 2.5600 & 2.5600 & 0.0117 & 0.0120& 23.50 & 4.5166 & 4.6383 & 0.0016 & 0.0016  \nl
18.25 & 2.7164 & 2.7164 & 0.0098 & 0.0100& 23.75 & 4.5768 & 4.7282 & 0.0015 & 0.0015  \nl 
18.50 & 2.8314 & 2.8314 & 0.0086 & 0.0088&       &        &        &        &         \nl
\enddata
\tablenotetext{a}{N and N$_{\rm Cor}$ are the differential galaxy counts before
and after correction for survey incompleteness, respectively. The units
are galaxies deg$^{-2}$ mag$^{-1}$.}
\tablenotetext{b}{The 1-sigma Poisson errors in log(N$_{\rm Cor})$.} 
\end{deluxetable}

\begin{deluxetable}{crrrrrrrrrrrrrrrrrr}
\tablewidth{0pt}
\scriptsize
\tablenum{2}
\tablecaption{Angular Two-Point Correlation Function \tablenotemark{a}}
\tablehead{
\colhead{$\theta$ } &
\colhead{$ I < 17 $} &
\colhead{$ 17 \le I < 18$} &
\colhead{$ 18 \le I < 19$} &
\colhead{$ 19 \le I < 20$} &
\colhead{$ 20 \le I < 21$} &
\colhead{$ 21 \le I < 21.5$} &
\colhead{$ 21.5 \le I < 22$} &
\colhead{$ 22 \le I < 22.5$} &
\colhead{$ 22.5 \le I < 23$}}
\startdata
0.37$'$& $ 192 \pm 84.5 $ & $ 136 \pm 33.8 $& $ 57.4 \pm 8.96 $& $ 29.4 \pm 3.02 $& $ 13.1 \pm 1.23 $& $ 8.56 \pm 1.36 $& $ 6.46 \pm 0.94 $& $ 6.87 \pm 0.98 $& $ 7.28 \pm 0.72 $ \nl
0.47 & $ 79.2 \pm 53.9 $ & $ 89.2 \pm 23.4 $& $ 48.3 \pm 6.94 $& $ 21.7 \pm 2.32 $& $ 12.2 \pm 0.98 $& $ 10.3 \pm 1.10 $& $ 9.44 \pm 0.77 $& $ 4.08 \pm 0.77 $& $ 3.96 \pm 0.57 $ \nl
0.59 & $ 131 \pm 46.8 $ & $ 74.0 \pm 17.8 $& $ 23.7 \pm 5.00 $& $ 19.2 \pm 1.85 $& $ 10.7 \pm 0.78 $& $ 7.86 \pm 0.86 $& $ 7.16 \pm 0.61 $& $ 4.31 \pm 0.62 $& $ 4.14 \pm 0.46 $ \nl
0.74 & $ 167 \pm 40.2 $ & $ 59.2 \pm 14.1 $& $ 27.2 \pm 4.08 $& $ 15.1 \pm 1.44 $& $ 10.6 \pm 0.62 $& $ 5.96 \pm 0.68 $& $ 5.90 \pm 0.48 $& $ 4.60 \pm 0.50 $& $ 4.39 \pm 0.37 $ \nl
0.94 & $ 146 \pm 29.7 $ & $ 63.5 \pm 11.3 $& $ 21.2 \pm 3.19 $& $ 16.2 \pm 1.17 $& $ 8.96 \pm 0.50 $& $ 6.16 \pm 0.55 $& $ 5.64 \pm 0.39 $& $ 4.90 \pm 0.41 $& $ 4.34 \pm 0.30 $ \nl
1.18 & $ 115 \pm 22.4 $ & $ 34.3 \pm 8.20 $& $ 16.5 \pm 2.51 $& $ 10.7 \pm 0.92 $& $ 7.18 \pm 0.40 $& $ 4.63 \pm 0.44 $& $ 4.61 \pm 0.31 $& $ 3.43 \pm 0.33 $& $ 4.04 \pm 0.25 $ \nl
1.48 & $ 95.7 \pm 16.9 $ & $ 44.7 \pm 6.76 $& $ 16.1 \pm 2.00 $& $ 10.4 \pm 0.74 $& $ 5.85 \pm 0.33 $& $ 4.97 \pm 0.35 $& $ 3.92 \pm 0.25 $& $ 3.99 \pm 0.27 $& $ 3.23 \pm 0.21 $ \nl
1.87 & $ 84.1 \pm 13.3 $ & $ 30.6 \pm 5.16 $& $ 13.4 \pm 1.59 $& $ 8.74 \pm 0.60 $& $ 5.19 \pm 0.28 $& $ 4.42 \pm 0.29 $& $ 3.52 \pm 0.21 $& $ 3.07 \pm 0.23 $& $ 2.99 \pm 0.17 $ \nl
2.35 & $ 67.1 \pm 10.2 $ & $ 29.0 \pm 4.12 $& $ 11.2 \pm 1.27 $& $ 6.70 \pm 0.50 $& $ 4.17 \pm 0.23 $& $ 2.76 \pm 0.23 $& $ 2.96 \pm 0.17 $& $ 2.83 \pm 0.19 $& $ 2.70 \pm 0.15 $ \nl
2.96 & $ 47.7 \pm 7.82 $ & $ 21.6 \pm 3.23 $& $ 11.2 \pm 1.03 $& $ 5.59 \pm 0.42 $& $ 3.91 \pm 0.20 $& $ 2.75 \pm 0.19 $& $ 2.51 \pm 0.15 $& $ 2.41 \pm 0.17 $& $ 2.48 \pm 0.13 $ \nl
3.72 & $ 54.8 \pm 6.50 $ & $ 26.1 \pm 2.66 $& $ 9.51 \pm 0.84 $& $ 4.59 \pm 0.36 $& $ 3.24 \pm 0.18 $& $ 2.47 \pm 0.16 $& $ 2.03 \pm 0.13 $& $ 2.39 \pm 0.15 $& $ 2.09 \pm 0.12 $ \nl
4.69 & $ 44.1 \pm 5.17 $ & $ 17.9 \pm 2.11 $& $ 7.72 \pm 0.69 $& $ 3.71 \pm 0.31 $& $ 2.20 \pm 0.17 $& $ 2.12 \pm 0.14 $& $ 1.80 \pm 0.11 $& $ 1.92 \pm 0.14 $& $ 1.76 \pm 0.11 $ \nl
5.90 & $ 27.1 \pm 4.06 $ & $ 17.6 \pm 1.75 $& $ 5.17 \pm 0.58 $& $ 3.25 \pm 0.28 $& $ 2.09 \pm 0.16 $& $ 1.53 \pm 0.13 $& $ 1.77 \pm 0.10 $& $ 1.53 \pm 0.13 $& $ 1.31 \pm 0.11 $ \nl
7.43 & $ 30.2 \pm 3.49 $ & $ 12.7 \pm 1.45 $& $ 4.89 \pm 0.50 $& $ 2.64 \pm 0.26 $& $ 1.66 \pm 0.15 $& $ 1.47 \pm 0.11 $& $ 1.28 \pm 0.10 $& $ 1.44 \pm 0.12 $& $ 1.21 \pm 0.10 $ \nl
9.35 & $ 17.7 \pm 2.93 $ & $ 8.81 \pm 1.23 $& $ 3.64 \pm 0.43 $& $ 2.22 \pm 0.25 $& $ 1.43 \pm 0.14 $& $ 1.20 \pm 0.11 $& $ 0.98 \pm 0.09 $& $ 1.23 \pm 0.12 $& $ 0.94 \pm 0.10 $ \nl
11.77 & $ 16.9 \pm 2.61 $ & $ 6.92 \pm 1.08 $& $ 2.94 \pm 0.39 $& $ 1.95 \pm 0.24 $& $ 1.28 \pm 0.14 $& $ 0.84 \pm 0.10 $& $ 0.75 \pm 0.09 $& $ 0.92 \pm 0.12 $& $ 0.71 \pm 0.10 $ \nl
14.82 & $ 10.3 \pm 2.36 $ & $ 5.22 \pm 0.97 $& $ 2.92 \pm 0.36 $& $ 1.53 \pm 0.23 $& $ 0.94 \pm 0.14 $& $ 0.60 \pm 0.10 $& $ 0.59 \pm 0.09 $& $ 0.69 \pm 0.11 $& $ 0.52 \pm 0.10 $ \nl
18.66 & $ 10.0 \pm 2.20 $ & $ 5.54 \pm 0.90 $& $ 1.81 \pm 0.34 $& $ 1.27 \pm 0.23 $& $ 0.75 \pm 0.14 $& $ 0.44 \pm 0.09 $& $ 0.40 \pm 0.09 $& $ 0.52 \pm 0.11 $& $ 0.36 \pm 0.10 $ \nl
23.49 & $ 11.1 \pm 2.10 $ & $ 5.72 \pm 0.86 $& $ 1.61 \pm 0.33 $& $ 0.95 \pm 0.22 $& $ 0.64 \pm 0.14 $& $ 0.45 \pm 0.09 $& $ 0.37 \pm 0.09 $& $ 0.53 \pm 0.11 $& $ 0.27 \pm 0.10 $ \nl
29.57 & $ 7.79 \pm 2.03 $ & $ 3.87 \pm 0.82 $& $ 0.91 \pm 0.32 $& $ 0.63 \pm 0.22 $& $ 0.38 \pm 0.14 $& $ 0.27 \pm 0.09 $& $ 0.26 \pm 0.09 $& $ 0.51 \pm 0.11 $& $ 0.22 \pm 0.10 $ \nl
37.22 & $ 7.51 \pm 1.99 $ & $ 2.60 \pm 0.80 $& $ 1.01 \pm 0.32 $& $ 0.57 \pm 0.22 $& $ 0.41 \pm 0.14 $& $ 0.23 \pm 0.09 $& $ 0.14 \pm 0.09 $& $ 0.35 \pm 0.11 $& $ 0.21 \pm 0.10 $ \nl
46.86 & $ 5.49 \pm 1.96 $ & $ 2.88 \pm 0.79 $& $ 0.91 \pm 0.31 $& $ 0.69 \pm 0.22 $& $ 0.35 \pm 0.14 $& $ 0.12 \pm 0.09 $& $ 0.05 \pm 0.09 $& $ 0.13 \pm 0.11 $& $ 0.25 \pm 0.10 $ \nl
58.99 & $ 2.21 \pm 1.94 $ & $ 2.37 \pm 0.78 $& $ 0.91 \pm 0.31 $& $ 0.67 \pm 0.22 $& $ 0.33 \pm 0.14 $& $ 0.12 \pm 0.09 $& $ -0.07 \pm 0.09 $& $ 0.23 \pm 0.11 $& $ 0.05 \pm 0.10 $ \nl
74.27 & $ 4.96 \pm 1.92 $ & $ 1.75 \pm 0.77 $& $ 0.67 \pm 0.31 $& $ 0.60 \pm 0.22 $& $ 0.23 \pm 0.14 $& $ 0.11 \pm 0.09 $& $ -0.06 \pm 0.09 $& $ 0.30 \pm 0.11 $& $ 0.06 \pm 0.10 $ \nl
93.50 & $ 5.94 \pm 1.92 $ & $ 0.79 \pm 0.77 $& $ 0.44 \pm 0.31 $& $ 0.53 \pm 0.22 $& $ 0.24 \pm 0.14 $& $ 0.23 \pm 0.09 $& $ 0.04 \pm 0.09 $& $ 0.40 \pm 0.11 $& $ 0.12 \pm 0.10 $ \nl
\enddata
\tablenotetext{a}{Divide $\omega(\theta)$ results by 100 to get actual values}
\end{deluxetable}

\begin{deluxetable}{ccccccccc}
\tablewidth{0pt}
\tablenum{3}
\tablecaption{Measured $\omega(\theta)$ Parameters}
\tablehead{
\colhead{$I_{median}$}  &
\colhead{N$_{\rm gal}$} &
\colhead{$\delta$\tablenotemark{a}}  &
\colhead{$\delta$\tablenotemark{b}}  &
\colhead{$\omega(0.5')_{Cor}$\tablenotemark{c}}  &
\colhead{$\omega(1')_{Cor}$\tablenotemark{c}}  &
\colhead{$\omega(3')_{Cor}$\tablenotemark{c}}  &
\colhead{IC Corr.} &
\colhead{$C_s$ Corr.\tablenotemark{d}}}
\startdata
16.46&  
1443               &
$-0.879 \pm 0.058$ &
$-0.688 \pm 0.087$ &
$202 \pm 101 $ &
$137 \pm 52.0 $ &
$56.4 \pm 15.6 $ &
0.0379 & 1.167\nl
17.64&  
3245               &
$-0.807 \pm 0.054$ &
$-0.739 \pm 0.079$ &
$80.4 \pm 41.2 $ &
$53.9 \pm 19.5 $ &
$25.0 \pm 6.46 $ &
0.0151 & 1.108\nl
18.62&  
9987               &
$-0.796 \pm 0.043$ &
$-0.690 \pm 0.061$ &
$30.1 \pm 11.9 $ &
$22.5 \pm 5.70 $ &
$9.86 \pm 2.06 $ &
0.0061 & 1.069\nl
19.61&  
27246              &
$-0.809 \pm 0.034$ &
$-0.808 \pm 0.041$ &
$23.7 \pm 4.17 $ &
$13.6 \pm 2.09 $ &
$5.56 \pm 0.83 $ &
0.0044 & 1.046\nl
20.59&  
62913              &
$-0.783 \pm 0.028$ &
$-0.739 \pm 0.030$ &
$13.6 \pm 1.75 $ &
$8.30 \pm 0.91 $ &
$3.55 \pm 0.41 $ &
0.0027 & 1.033\nl
21.27&  
55245              &
$-0.788 \pm 0.033$ &
$-0.639 \pm 0.041$ &
$8.91 \pm 1.96 $ &
$6.19 \pm 0.99 $ &
$2.78 \pm 0.39 $ &
0.0018 & 1.025\nl
21.77&  
78082              &
$-0.742 \pm 0.029$ &
$-0.674 \pm 0.034$ &
$8.22 \pm 1.37 $ &
$5.26 \pm 0.70 $ &
$2.48 \pm 0.39 $ &
0.0017 & 1.022\nl
22.27&  
106975             &
$-0.627 \pm 0.034$ &
$-0.475 \pm 0.042$ &
$5.89 \pm 1.39 $ &
$4.55 \pm 0.74 $ &
$2.44 \pm 0.34 $ &
0.0014 & 1.020\nl
22.77&  
137538             &
$-0.697 \pm 0.034$ &
$-0.483 \pm 0.035$ &
$4.05 \pm 1.03 $ &
$4.19 \pm 0.55 $ &
$2.25 \pm 0.27 $ &
0.0016 & 1.019\nl
\enddata

\tablenotetext{a}{Slope for the best fit power law in the range 
$1' \le \theta \le 20'$.}

\tablenotetext{b}{Slope for the best fit power law in the range 
$0.5' \le \theta \le 5'$.}

\tablenotetext{c}{Divide by 100 to get actual amplitude.
Based on a local interpolation to provide a more accurate
estimate of the amplitude. The IC and stellar contamination ($C_s$) corrections
have been applied. Corrected 
$\omega(\theta) = C_s (\omega(\theta)_{obs} + IC)$}

\tablenotetext{d}{Stellar contamination correction 
factor,~$C_s = N_{Obj}^2/(N_{Obj}-N_{Star})^2$.}

\end{deluxetable}

\begin{deluxetable}{ccccccccccc}
\tablewidth{0pt}
\small
\tablenum{4}
\tablecaption{Best-fit $r_o$ and $\epsilon$ values (1 arcminute scale)}
\tablehead{
\colhead{ } &
\colhead{ } &
\colhead{$r_o$ } &
\colhead{$\sigma_{r_o}$ at} &
\colhead{ } &
\colhead{$\sigma_{\epsilon}$\tablenotemark{a}\ at} &
\colhead{ } &
\colhead{Covar.} &
\colhead{Covar.} &
\colhead{ } &
\colhead{ } \\
\colhead{Sample} &
\colhead{$N(z)$ Model} &
\colhead{($h^{-1}$ Mpc)}  &
\colhead{fixed $\epsilon$} &
\colhead{$\epsilon$} &
\colhead{fixed $r_o$} &
\colhead{$\chi^2_{\nu}$\tablenotemark{b} } &
\colhead{slope\tablenotemark{c}} &
\colhead{Intercept\tablenotemark{c}} &
\colhead{$r_o$ Range\tablenotemark{d}} &
\colhead{$\epsilon$ Range\tablenotemark{d}} 
}
\startdata
$I \le 20$& CFRS             &$4.70$ & 0.32&$-1.50$& 0.42& 0.11& 0.768&5.822& 3 to 7& $-3$ to $+1$\\
$I \le 20$& EvLF $\beta=1$   &$5.30$ & 0.35&$+2.30$& 0.64& 0.29& 0.555&4.577& 3 to 7& $-3$ to $+4$\\
$I \le 20$& EvLF $\beta=1.5$ &$5.20$ & 0.33&$+1.20$& 0.66& 0.30& 0.495&4.209& 3 to 7& $-3$ to $+4$\\
$I \le 20$& EvLF $\beta=2$   &$5.30$ & 0.36&$+0.40$& 0.57& 0.30& 0.624&5.060& 3 to 7& $-3$ to $+4$\\
   &   &   &  &  &  &   &   &   &  \\
$I \le 23$& CFRS             &$5.60$ & 0.23&$-0.20$& 0.17& 0.24& 1.356&5.874& 4 to 7& $-1.5$ to $+1$\\
$I \le 23$& EvLF $\beta=1$   &$3.80$ & 0.15&$-0.80$& 0.21& 0.39& 0.713&4.389& 2.7 to 5& $-2.5$ to $+1$\\
$I \le 23$& EvLF $\beta=1.5$ &$3.80$ & 0.15&$-1.40$& 0.18& 0.35& 0.830&4.954& 2.8 to 5.1& $-2.7$ to 0\\
$I \le 23$& EvLF $\beta=2$   &$4.10$ & 0.16&$-1.70$& 0.16& 0.26& 1.025&5.798& 3 to 5.2& $-2.8$ to $-0.5$\\
   &   &   &  &  &  &   &   &   &  \\
$I \le 23$& CFRS             &$5.50$ Fixed& \nodata&$-0.30$& 0.75& 0.25&\nodata&\nodata &\nodata &\nodata\\
$I \le 23$& EvLF $\beta=1$   &$5.50$ Fixed& \nodata&$+1.30$& 0.90& 2.09&\nodata&\nodata &\nodata &\nodata\\
$I \le 23$& EvLF $\beta=1.5$ &$5.50$ Fixed& \nodata&$+0.50$& 0.80& 2.06&\nodata&\nodata &\nodata &\nodata\\
$I \le 23$& EvLF $\beta=2$   &$5.50$ Fixed& \nodata&$-0.40$& 0.70& 1.83&\nodata&\nodata &\nodata &\nodata\\
\enddata
\tablenotetext{a}{Uncertainties in $\epsilon$ at fixed $r_o$ are the $1\sigma$
values except when $r_o$ is kept fixed. In that case, the errors shown are
the $2\sigma$ values.}
\tablenotetext{b}{The reduced $\chi^2$ value.} 
\tablenotetext{c}{The slope and intercept for the best-fit
line to the $1\sigma$ $\chi^2$ contour. 
The line is $r_o = ({\rm slope})\epsilon\ +\ {\rm intercept}$. Values
are the average parameters derived from
the fits $r_o = f(\epsilon)$ and $\epsilon = f(r_o)$.}
\tablenotetext{d}{The $r_o$ and $\epsilon$ range over which the fit 
to the $1\sigma$ $\chi^2$ contour is valid.} 
\end{deluxetable}
\end{document}